\documentclass[%
twocolumn,
superscriptaddress,
 amsmath,amssymb,
 aps,
prc,
floatfix,
]{revtex4-1}

\usepackage{amsmath}
\usepackage{amssymb}
\usepackage{amsfonts}

\usepackage{xcolor}

\usepackage{dsfont}

\usepackage{braket}
\usepackage{graphicx}
\usepackage{dcolumn}
\usepackage{bm}

\newcommand{\WignerSIXj}[6]
	{
	\left\{
		\begin{array}{ccc}
			#1 & #2 & #3 \\
   			#4 & #5 & #6
		\end{array}
	\right\}
	}

\newcommand{\WignerNINEj}[9]
	{
	\left\{
		\begin{array}{ccc}
			#1 & #2 & #3 \\
   			#4 & #5 & #6 \\
   			#7 & #8 & #9
		\end{array}
	\right\}
	}

\newcommand{\UNINEj}[9]
	{
	  \text{ U} \! \left[
		\begin{array}{ccc}
			#1 & #2 & #3 \\
   			#4 & #5 & #6 \\
   			#7 & #8 & #9
		\end{array}
	\right]
	}

\newcommand{\SuThreeCGReduced}[7]
	{
	  { \braket{ #1 #2 ; #3 #4 || #5 #6 } }_{ #7 }
	}

\newcommand{\half}[0]
	{
	  \frac{ 1 }{ 2 }
	}

\newcommand{\SuThreeUNINEFirst}[0]
	{
	  \text{ U} \! \left[
		\begin{array}{cccc}
		  ({ \eta }_{ b }\, 0) & { \omega }_{0} & ({ \eta }_{ a }\, 0) & 1 \\
		  (\eta\, 0) & { \omega }_{\rm p} & (\eta' \, 0) & 1 \\
		  { {\omega} }_{ b } & {\omega}_{0} '& { {\omega} }_{ a } & \rho_{a} \\
		  1 & { \rho }_{0}' & 1 &  -
		\end{array}
	\right]
	}

\newcommand{\SuThreeUNINESecond}[0]
	{
	  \text{ U} \! \left[
		\begin{array}{cccc}
		  { \omega }_{ 1 } & { \omega }_{0} & { \omega }_{ 1 }' & { \rho }_{ 0 } \\
		  (\eta \, 0) & { \omega }_{\rm p} & (\eta' \, 0) & 1 \\
		  \omega & {\omega}_{0} '& \omega' & \rho' \\
		  1 & { \rho }_{0}' & 1 &  -
		\end{array}
	\right]
	}

\newcommand{\SU}[1]{\ensuremath{\mathrm{SU}( #1 )}}

\newcommand{\SpR}[1]{\ensuremath{\mathrm{Sp}( #1,\mathbb{R} )}}
\newcommand{\hw}{\ensuremath{\hbar\Omega}}

\begin{document}

\preprint{APS/123-QED}

\title{Efficacy of the symmetry-adapted basis for \textit{ab initio} nucleon-nucleus interactions \\ for light- and intermediate-mass nuclei } 

\author{A. Mercenne}
\affiliation{Department of Physics and Astronomy, Louisiana State University, Baton Rouge, LA 70803, USA}
\affiliation{Center for Theoretical Physics, Sloane Physics Laboratory, Yale University, New Haven, Connecticut 06520, USA}
\author{K. D. Launey}%
\affiliation{Department of Physics and Astronomy, Louisiana State University, Baton Rouge, LA 70803, USA}
\author{T. Dytrych}
\affiliation{Nuclear Physics Institute, Academy of Sciences of the Czech Republic, 25068 {\v{R}}e{\v{z}}, Czech Republic}%
\author{J. E. Escher}
\affiliation{Lawrence Livermore National Laboratory, Livermore, CA, 94550, USA}%
\author{S. Quaglioni }
\affiliation{Lawrence Livermore National Laboratory, Livermore, CA, 94550, USA}%
\author{G. H. Sargsyan}%
\affiliation{Department of Physics and Astronomy, Louisiana State University, Baton Rouge, LA 70803, USA}%
\author{J. P. Draayer}
\affiliation{Department of Physics and Astronomy, Louisiana State University, Baton Rouge, LA 70803, USA}

\date{\today}

\begin{abstract}
We study the efficacy of a new \textit{ab initio} framework that combines the symmetry-adapted (SA) no-core shell-model approach with the resonating group method (RGM) for unified descriptions of nuclear structure and reactions. We obtain \textit{ab initio} neutron-nucleus interactions for $^{4}$He, $^{16}$O, and $^{20}$Ne targets, starting with realistic nucleon-nucleon potentials. We discuss the effect of increasing  model space sizes and symmetry-based selections on the SA-RGM norm and direct potential kernels, as well as on phase shifts, which are the input to calculations of cross sections. We demonstrate the efficacy of the SA basis and its scalability with particle numbers and model space dimensions, with a view toward \textit{ab initio} descriptions of nucleon scattering and capture reactions up through the medium-mass region.
\end{abstract}

\pacs{Valid PACS appear here}% PACS, the Physics and Astronomy
\maketitle

\section{Introduction}
{

{\it Ab initio} descriptions of spherical and deformed nuclei up through the calcium region are now possible within a no-core shell-model framework, by utilizing emerging symplectic symmetry in nuclei.  In particular, the symmetry-adapted no-core shell-model (SA-NCSM) \cite{LauneyDD16,DytrychLDRWRBB20} uses a physically relevant symmetry-adapted (SA) basis that can achieve significantly reduced  model spaces compared to the corresponding complete ultra-large  model spaces, without compromising the accuracy of results for various observables \cite{DytrychHLDMVLO14,LauneyDD16,BakerLBND20}.  This enables the SA-NCSM to accommodate contributions from more shells and to describe heavier nuclei, such as $^{20}$Ne \cite{DytrychLDRWRBB20}, $^{21}$Mg \cite{Ruotsalainen19}, $^{22}$Mg \cite{Henderson:2017dqc}, $^{28}$Mg \cite{PhysRevC.100.014322},  as well as $^{32}$Ne and $^{48}$Ti \cite{LauneySOTANCP42018,LauneyMD_ARNPS21}.
The access to higher-lying shells makes the SA basis suitable for describing nuclear reactions \cite{LauneyMD_ARNPS21}, the processes that are typically studied in experiments and govern stellar evolution. Remarkable progress has been made in first-principle descriptions to  scattering and nuclear reactions for light nuclei (for an overview, see \cite{FRIBTAwhite2018,doi:10.1080/10619127.2020.1752089}), 
including studies of elastic scattering \cite{NollettPWCH07,HagenDHP07,PhysRevLett.101.092501,ElhatisariLRE15,PhysRevLett.125.112503,mercenne_2019,BurrowsBEWLMP20}, photoabsorption \cite{PhysRevC.90.064619}, transfer \cite{NavratilQ12} and capture reactions \cite{PhysRevLett.105.232502}, $\alpha$ widths \cite{kravvaris_2017,DreyfussLESBDD20} and resonant states \cite{lazauskas_2019}, as well as thermonuclear fusion \cite{HupinQN19}. In this paper, we show that expanding the reach of \textit{ab initio} reactions to deformed and heavier targets is now feasible with the SA basis.

Microscopic approaches to nuclear reactions take into account nucleon degrees of freedom along with their correlations within and among the reaction fragments. Coupled with realistic inter-nucleon interactions, such as the ones derived in the framework of chiral effective field theory \cite{BedaqueVKolck02,EpelbaumNGKMW02,EntemM03,Epelbaum06}, these approaches  provide \textit{ab initio} calculations of reaction observables. 
One of the  earliest and very successful microscopic approaches to nuclear reactions is the resonating-group method (RGM)  \cite{WildermuthT77,RGM_tang_1978}.  In the RGM, nucleons are organized within different groups, or clusters, ``resonating'' through the inter-cluster exchange of nucleons. Most importantly, the cluster system is translationally invariant, and the Pauli exclusion principle is enforced by the antisymmetrization between the different clusters. 
All of these features make this method particularly suitable for unified descriptions of nuclear structure and reaction observables. 
Following the success of the Elliott model \cite{Elliott58,Elliott58b}, showing that a leading (most deformed) \SU{3} shell-model configuration describes reasonably well the ground-state rotational band in intermediate-mass nuclei,  the RGM has been extensively used with an \SU{3} basis and its no-core shell-model extension, the symplectic \SpR{3} basis
\cite{Hecht77_NPA283,Hecht_1978,HechtS82}.  
Applications of the model with Gaussian interactions have successfully calculated  $\alpha$ and $^8$Be cluster amplitudes, spectroscopic amplitudes for heavy-fragment clusters, and sub-Coulomb $^{12}$C+$^{12}$C resonances \cite{HechtRSZ81,HechtB82,SuzukiH82}. The formalism has been extended by utilizing a mixed no-core shell-model \SpR{3} plus RGM cluster basis \cite{Suzuki86,SuzukiH86, SuzukiH87}, and applied to studies of the monopole and quadrupole strengths in light nuclei~\cite{Suzuki87, SuzukiH89}, as well as the $\alpha$+$^{12}$C cluster system \cite{Suzuki76a,Suzuki76b,SuzukiLYV03}.

More recently, a successful first-principle description of scattering and reactions has been realized by implementing the RGM using \textit{ab initio} NCSM \cite{NavratilQSB09,BarrettNV13} wave functions  for the clusters in a formalism known as NCSM/RGM \cite{quaglioni_2008,QuaglioniN09,PhysRevC.88.054622}
and, later, by fully combining  the two  approaches
into the generalized \textit{ab initio} cluster expansion of the no-core shell model with continuum
(NCSMC) \cite{BaroniNQ13,PhysRevC.87.034326}. 
These methods, which have enabled predictions of  nucleon \cite{PhysRevC.90.061601,PhysRevLett.117.242501}, deuteron \cite{PhysRevLett.114.212502} and alpha \cite{KravvarisQHN2020} scattering off light targets, as well as polarized deuterium-tritium  fusion \cite{HupinQN19} from chiral nucleon-nucleon (NN) and three—nucleon (3N) forces, are reviewed in Refs. \cite{1402-4896-91-5-053002,doi:10.1080/10619127.2020.1752089}. 

In addition, the Gamow shell model coupled-channel approach combines the  RGM with a continuum core-valence shell approach and allows for descriptions of nuclear reactions of heavier systems \cite{jaganathen_2014,fossez_2015,mercenne_2019}.

The goal of this paper is to show the efficacy of a new approach that can extend the study of \textit{ab initio} reactions to medium mass nuclei by using the SA-NCSM approach \cite{LauneyDD16,DytrychLDRWRBB20}.
The SA framework takes advantage of symmetries inherent to nuclei and of group theoretical algorithms, and reorganizes the model space into a physically relevant basis. This allows us to account for the relevant correlations within only a few dominant components and, hence, achieve manageable Hamiltonian matrix sizes. 
In this paper, we present a new formalism of the RGM, one that admits the use of the SA basis, and we demonstrate the capability and potential of the approach for light and intermediate-mass nuclei. 
The formalism of the SA-RGM framework is presented in Sec. \ref{Formalism}, where we discuss RGM kernels computed using the SA basis. The sensitivity of the kernels on different selected model spaces and model space sizes is discussed in Sec. \ref{SAval} for a $^4$He target and in Sec. \ref{SAappl} for intermediate-mass $^{16}$O and $^{20}$Ne targets. Section \ref{dimension_analysis} presents an analysis of the basis dimension and its scaling with model space sizes and particle numbers. Finally, Sec. \ref{concl} outlines the conclusions.

}

\section{Theoretical framework}
\label{Formalism}
{
Traditionally, the RGM adopts microscopic cluster wave functions as basis functions to describe the motion of a system of two or more clusters (see, e.g., Refs. \cite{QuaglioniN09,PhysRevC.88.034320}). 
We consider two nuclear fragments, or binary-cluster nuclear reactions.
For two clusters $A$ and $a$, the cluster states for a channel $c$ are defined as:
  \begin{eqnarray}
   \ket{ { \Phi }_{ c r }^{J^\pi M} }=&& { \{ \{{ \ket{ (A) \mathfrak{a}_1 I_1^{\pi_1}} \times \ket{ (a) \mathfrak{a}_2 I_2^{\pi_2}} } \}^I \times Y_{ \ell } ({ \hat{ r } }_{ A,a }) \}^{J^\pi M} } 
  \nonumber \\
  &&\times \frac{ \delta(r - { r }_{ A,a }) }{ r { r }_{ A,a } } ,
    \end{eqnarray}
    where  the cluster system is defined for a channel $c  = \{  \mathfrak{a}_1, I_1^{\pi_1}, \mathfrak{a}_2, I_2^{\pi_2},I,\ell\}$, which is labeled by the angular momentum  (spin) and parity of each of the clusters  and the total spin of the clusters $I$ (the labels  $\mathfrak{a}_1$ and $\mathfrak{a}_2$ denote all other quantum numbers needed to fully characterize their respective states), and the orbital momentum $l$. 
For particle laboratory coordinates $\vec r_i$ (used in this study), the separation distance  between the center-of-mass of the two clusters is determined from ${ \vec r }_{ A,a }={1 \over A}\sum_{i=1}^A \vec r_i -{1 \over a}\sum_{i=A+1}^{A+a} \vec r_i$. The distance $r$ between the clusters defines the cluster states and the RGM kernels, as shown below, and as an integration variable facilitates the treatment of the inter-cluster antisymmetrization. Namely,  the $A+a$ nuclear wave function is expressed in terms of the cluster states as
  \begin{equation}
    \ket{ { \Psi } ^{J^\pi}} = \sum_{c} \int_{r} dr { r }^{ 2 } \frac{ g^{J^\pi} _{ c }(r) }{ r } { \mathcal{A}_c } \ket{ { \Phi }_{ c r }^{J^\pi} } \;,
    \label{RGM_ansatz}
  \end{equation}
with unknown amplitudes ${ { g }_{ c }^{J^\pi}(r) }$ that are determined  by solving the integral Hill-Wheeler equations for a given total energy $E$ in the $A+a$ center-of-mass frame: 
  \begin{equation}
    \sum_{c} \int dr { r }^{ 2 } \left[ { H }_{ c' c } (r',r) - E { N }_{ c'c }(r',r) \right] \frac{ { g }_{ c }^{J^\pi}(r) }{ r } = 0.
    \label{RGM_equations}
  \end{equation}
  Here,  $H _{ c'c }(r',r) ={ \bra{ { \Phi }_{ c' r' } ^{J^\pi}} { \mathcal{A} }_{c'} \hat H { \mathcal{A} }_{c} \ket{ { \Phi }_{ c r }^{J^\pi} } }$ is the Hamiltonian kernel and $N_{ c' c }(r',r)={ \bra{ { \Phi }_{ c' r' }^{J^\pi} } { \mathcal{A} } _{c'} { \mathcal{A} }_c \ket{ { \Phi }_{ c r }^{J^\pi} } }$ is the norm kernel, where ${ { \mathcal{A} } }$ is the inter-cluster antisymmetrizer. The kernels are computed by using the microscopic wave functions of the clusters that can be obtained in the \textit{ab initio} NCSM and SA-NCSM. Once the kernels are computed, Eq. (\ref{RGM_equations}) can then be solved using the microscopic ${ R }$-matrix approach \cite{DescouvemontB10,Descouvemont16}.
 
In the SA-RGM,  the target nucleus of $A$ particles is described by SA-NCSM many-body wave functions. In the SA-NCSM, the many-body basis is labeled by irreducible representations (irreps) according to the group chain \cite{Elliott58,Elliott58b}: 
  \begin{equation}
    \left[ {\text{SU}(3)}_{ (\lambda\, \mu) } \underset{\kappa}{\supset} {\text{SO}(3)}_{ L } \supset {\text{SO}(2)}_{ { M }_{ L } } \right] \otimes \left[ \SU{2}_S \supset \SU{1}_{M_S} \right] .
    \label{SU3SO3SO2GroupChain}
  \end{equation}
 The ${ (\lambda\, \mu) }$ quantum numbers  label an \SU{3} irrep and can be related to the average deformation through the established link with the well-known  parameters, deformation  ${ \beta }$ and triaxiality ${ \gamma }$ \cite{CastanosDL88,MustonenGAB2018}.
  The label ${ \kappa }$ distinguishes multiple occurrences of the same orbital momentum ${ L }$ in the parent irrep ${ (\lambda\, \mu) }$, and ${ M }_{ L } $ is the projection. These quantum numbers define the spatial degrees of freedom, which can then be coupled to the intrinsic spin ($S$) to yield a good total angular momentum.

 Specifically, a target state with total angular momentum and parity ${ I_1 }^{\pi_1}$ (and projection ${ M_1 }$)  is constructed in terms of the SA basis:
  \begin{equation}
    \ket{ (A) \mathfrak{a}_1 I_1^{\pi_1} M_1 } = \!\!\!
   \sum_{ \substack{ { \mathfrak{b} }_1  \omega_1 \\ \kappa_1L_1S_1 }} { C }_{ { \mathfrak{b} }_1 }^{ { \omega }_{ 1 } { \kappa }_{ 1 } L_1{ S }_{ 1 } I_1}  \ket{ { \mathfrak{b} }_1  \omega_1 \kappa_1 (L_1S_1)  I_1^{\pi_1} M_1 },
    \label{SU3wf}
  \end{equation}
  where the labels are defined as 
  $\mathfrak{b} \equiv  \left\{ \dots   {\omega}_{ \rm p } { \omega}_{\rm n } \rho N; { S }_{ \rm p } { S }_{\rm n } \right\}$  
  and deformation $\omega \equiv (\lambda\, \mu) $ (it is understood that the coefficients $C$ are for given ${\pi_1}$, which is omitted from labeling). Protons and neutrons are labeled by p and n, respectively,  and $S$ labels the intrinsic spin (``$\dots$" denotes all additional quantum numbers including $\mathfrak{a}_1$). The \SU{3} outer multiplicity ${ \rho }$ \cite{DraayerLPL89} results from the coupling of the proton deformation with that of neutrons to total deformation $\omega_1 $. ${ N }$ labels the total HO excitations above the valence-shell configuration and is truncated at a maximum value ($N\leq N_{\rm max}$), which determines the model space size.  
  
  For a single-particle projectile, the SA-RGM channel basis states can thus be defined for a channel $\{{\nu_1;\nu}\}=\{{ \omega }_{ 1 } { \kappa }_{ 1 } (L_1{ S }_{ 1 }); \omega \kappa (L S)\}$ [related to channel $c$ in Eq. (\ref{nutoc})] as:
  \begin{equation}
    \ket{ { \Phi }_{ \nu_1 I_1; \eta }^{ \nu J^\pi M } } = \sum_{ { \mathfrak{b} }_{ 1 } } 
    { C }_{ { \mathfrak{b} }_{ 1 } }^{ \nu_1 I_1} 
    { \left\{ \ket{ { \mathfrak{b} }_{ 1 } { \omega }_{ 1 } { S }_{ 1 } } \times \ket{ (\eta \, 0) {1 \over 2} } \right\} }^{ \nu J M },
    \label{SU3RGMstates}
  \end{equation}
 where the \SU{3} basis states for the target  are coupled  to the HO single-particle states of the projectile with $(\eta \,0)$ SU(3) quantum numbers and spin $ {1 \over 2}$ (we will omit the parity $\pi$ from the notation throughout the paper for simplicity). We note that the SU(3) outer multiplicity associated with the coupling of $\omega_1$ and $(\eta \,0)$ is 1, and hence, omitted from the notations.
 Remarkably, there is no dependence on the orbital momentum of the projectile, only on the shell number it occupies, $\eta$. Furthermore, the summation over 
 ${ { \mathfrak{b} }_{ 1} }$ 
 implies that the SA-RGM basis requires only a part of the information present in the SA basis. 
 
 The SA-RGM basis is used to calculate the RGM kernels, which is the main computational task in RGM  \cite{QuaglioniN09}. These include the norm kernel, which is the overlap between antisymmetrized non-orthogonal RGM basis states. It consists of a direct part (a Dirac delta function), which dominates at large relative distances, and an exchange part that takes into account the Pauli principle at short distances. The exchange norm kernel is related to the permutation operator ${ { { P } }_{A,A+1 } }$ that exchanges the nucleon projectile with another nucleon within the target, thereby ensuring antisymmetrization (cf. \cite{QuaglioniN09}):
  \begin{align}
    { N }_{ c'c }^{ \text{ex} } (r',r) & = - \bra{ { \Phi }_{ c'r' }^{ JM } } \sum_{i=1}^{A} { \hat{ P } }_{ i,A+1 } \ket{ { \Phi }_{ c r }^{ JM } } \nonumber \\
    & = - A \bra{ { \Phi }_{ c'r' }^{ JM } } { \hat{ P } }_{ A,A+1 } \ket{ { \Phi }_{ c r }^{ JM } }
    \label{ExNKernel}
  \end{align}
 
 The exchange norm kernel in the SA-RGM basis is thus reduced to evaluating the following (similarly, for the Hamiltonian kernels):
 \begin{widetext}
  \begin{eqnarray}
     \bra{ { \Phi }_{\nu_1' I_1'; \eta'  }^{ \nu' JM } } { { P } }_{ A,A+1 } \ket{ { \Phi }_{ \nu_1 I_1; \eta }^{ \nu JM } } 
     &=& { \delta }_{ \nu' \nu } \sum_{ { \omega }_{0} { S }_{0} { \rho }_{ 0} } { \Pi }_{ { S }_{0} { S }_{ 1 }' } { (-1) }^{ \eta + \eta' - { \omega }_{0} } { (-1) }^{ { S }_{ 1 } + \frac{ 1 }{ 2 } + S' } 
     \WignerSIXj{ { S }_{ 1 } }{ { S }_{0} }{ { S }_{ 1 }' }{ \frac{ 1 }{ 2 } }{ S }{ \frac{ 1 }{ 2 } } 
     \nonumber \\
   & \times &  \sqrt{ \frac{ {\rm dim} {\, \omega }_{0} }{ {\rm dim}\, (\eta\,0) } }   
    U\left[  { \omega }_{ 1 } { \omega }_{0}  \omega (\eta\,'0) ; { \omega}_{ 1 }' { \rho }_{ 0} 1 (\eta\,0) 1 1 \right] 
    { {\rho} }_{ \eta \eta' }^{  { \rho }_{ 0} { \omega }_{0} { S }_{0}} \left( \nu_1' I_1' ; \nu_1 I_1 \right),
    \label{ExchangeMatrixSU3}
  \end{eqnarray}
  where ${ U\left[  \dots  \right] }$ is the \SU{3} 6-$(\lambda\,\mu)$ recoupling coefficient \cite{DraayerSU3_1}, analogous to the SU(2) 6-$j$ symbol, $\dim\, (\lambda\,\mu)={1\over 2}(\lambda+1)(\mu+1)(\lambda+\mu+2)$, and the \SU{3} one-body density matrix elements are  defined as:
\begin{equation}
   { {\rho} }_{ \eta  \eta' }^{  { \rho }_{ 0} { \omega }_{0} { S }_{0} } \left( \nu_1' I_1' ; \nu_1 I_1\right) = \sum_{ { \mathfrak{b} }_{ 1 } { \mathfrak{b} }_{ 1 }' } { C }_{ { \mathfrak{b} }_{ 1 }' }^{ \nu_1' I_1'} { C }_{ { \mathfrak{b} }_{ 1 } }^{ \nu_1 I_1}  \langle { { \mathfrak{b} }_{ 1 }' { \omega }_{ 1 }' { S }_{ 1 }' } ||| { \{ { a }_{ (\eta \,0) \frac{ 1 }{ 2 } }^{ \dagger } \times { \tilde a }_{ {(0\, \eta')}\frac{ 1 }{ 2 } } \} }^{ { \omega }_{0} { S }_{0} } ||| { { { \mathfrak{b} }_{ 1 } { \omega }_{ 1 } { S }_{ 1 } } } \rangle_{ { \rho }_{ 0} },
  \label{O_objects}
\end{equation}
 \end{widetext}
where $a^\dagger_{(\eta\, 0) lm_l \half m_s} \equiv a^\dagger_{\eta lm_l \half m_s}$ and $a_{\eta lm_l \half m_s}$ creates and annihilates, respectively, a particle of spin 1/2 in the $\eta$-th HO shell, and $\tilde a_{(0 \,\eta) l -m_l \half -m_s} \equiv (-1)^{\eta +l -m_l +s -m_s} a_{\eta lm_l \half m_s}$ is the annihilation SU(3) tensor operator.  
The matrix elements of the $ \rho$ density can be quickly computed on the fly in the SA basis. Their computation can utilize an efficacious algorithm that exploits organization 
of SA basis states in terms of subspaces of SU(3) irreps and the factorization of spatial SU(3) and SU(2) spin degrees of freedom~\cite{DytrychSBDV_PRL07,LauneyDD16},
and this can be done prior to the computation of the kernels. It is notable that, as a result of the Kronecker delta function $\delta_{\nu \nu'}$ in Eq. (\ref{ExchangeMatrixSU3}), the exchange part of the norm kernel turns out to be block-diagonal in the SA-RGM basis.
The reason is that the operator ${ { { P } }_{  } }$ is an SU(3) scalar and spin scalar, and therefore  preserves deformation and spin of the composite $A+1$ system (note that it may change the $\omega_1$ deformation of the target itself).

%This procedure 
Eq. (\ref{ExchangeMatrixSU3}) allows the kernels to be calculated, for  each ${ J ^\pi M }$, through the SA-RGM channel basis of Eq. (\ref{SU3RGMstates}) that only depends on  the  deformation, rotation, and spin of the target $\nu_1$ (that is, ${ \omega }_{ 1 } { \kappa }_{ 1 }L_1 { S }_{ 1 }$), and the deformation, rotation, and spin of the target-projectile system $\nu$ (that is, $\omega \kappa LS $). From this, it is clear that the SA offers two main advantages: first, calculations utilize group-theoretical algorithms that use a reduced subset of quantum numbers $\nu$ and $\nu_1$, and second, the number of \SU{3} configurations in the target wave function, we find, is a manageable number when compared to the complete model-space size. This results in a manageable number of  configurations for the target-projectile system based on \SU{3} and SU(2) selection rules, namely, $\omega=\omega_1 \times (\eta \, 0)$ and $S=S_1 \times {1\over 2}$ (for further details on scalability, see Sec. \ref{dimension_analysis}). 

Another advantage of the SA scheme is that the dependence on the  orbital momentum $\ell$ is recovered in the very last step:
  \begin{align}
    & \ket{ { \Phi }_{ c r }^{ JM } } = \sum_{ \eta } { R }_{ \eta \ell }(r) \sum_{j} { \Pi }_{ sj } { (-1) }^{ { I }_{ 1 } + J + j } \nonumber \\
    & \times \WignerSIXj{ { I }_{ 1 } }{ \half }{ s }{ \ell }{ J }{ j } \sum_{ \substack{\nu \\ \nu_1 } } { \Pi }_{ LS { I }_{ 1 }j } \SuThreeCGReduced{ { \omega }_{ 1 } }{ { \kappa }_{ 1 } {L}_{1} }{ (\eta\,0) }{ \ell }{ \omega }{ \kappa L }{  } \nonumber \\
    & \times   \WignerNINEj{ { L }_{ 1 } }{ { S }_{ 1 } }{ { I }_{ 1 } }{ \ell }{ \half }{ j }{ L }{ S }{ J } \ket{ { \Phi }_{ \nu_1 I_1 ; \eta }^{ \nu JM } }.
    \label{nutoc}
  \end{align}
This wave function is then used in a microscopic ${ R }$-matrix approach  \cite{DescouvemontB10} to calculate phase shifts and cross sections. 

To study the efficacy of the SA scheme, we focus on  the norm and potential kernels. For the potential kernel, we consider only the part that involves the projectile and a single nucleon in the target (similarly to Ref. \cite{BurrowsBEWLMP20}), that is, the  potential kernel of particle-rank one, denoted here as ${ V }_{c'c}^{ (1) }(r',r)$ (cf. \cite{QuaglioniN09}):
  \begin{align}
    {V}_{c'c}^{(1)}(r',r) 
    & \equiv A \bra{ { \Phi }_{ c'r' }^{ JM } } { \hat{ V } }_{ A,A+1 }(1-{ \hat{ P } }_{ A,A+1 }) \ket{ { \Phi }_{ c r }^{ JM } }.
    \label{dHKernel}
  \end{align}
Note that the exchange of two nucleons that interact with each other is part of this kernel.  We
do not consider the particle-rank two potential kernel that accounts for the projectile exchanging with one nucleon in the target and interacting  with another nucleon (called exchange potential kernel in \cite{QuaglioniN09}). Since the goal of this study is to validate the use of the SA scheme against the use of complete model spaces, we expect that the particle-rank two potential kernel will benefit from advantages similar to those shown in the next section. The reason is that the main advantage stems from the reductions of the number of basis states needed to describe the target wavefunctions. Such a reduction ensures that one-body densities, along with the two-body densities that will be needed for the particle-rank two potential kernels, are computed for wavefunctions that span only a fraction of the complete model space (as discussed in Sec. \ref{dimension_analysis}).

The derivation of the potential kernel in the SA-RGM basis follows a  procedure similar to that for the norm kernel:
\begin{widetext}
  \begin{align}
    & \bra{ { \Phi }_{ {\nu}_{1}' {I}_{1}'; \eta ' }^{ \nu' JM } } ({ \hat{ V } }_{ A,A+1 }(1 - { \hat{ P } }_{ A,A+1 } )) \ket{ { \Phi }_{ \nu_1 {I}_{1}; \eta }^{ \nu JM } } \nonumber \\
    & = \sum_{ \substack{ { S }_{ b } { S }_{ a } \\ { S }_{0} { S }_{\rm p} } } { \left( \frac{ { \Pi }_{ { S }_{\rm p} { S }_{ a } } }{ { \Pi }_{ {S}_{0}' \half } } \right) }^{ 2 } \frac{ { \Pi }_{ { S }_{0} } }{ { \Pi }_{ \half } } \frac{ 1 }{ { \Pi }_{ { S }_{ 1 }' { S }_{ a } } }  \UNINEj{ \half }{ { S }_{0} }{ \half }{ \half }{ { S }_{\rm p} }{ \half }{ { S }_{ b } }{ S_{0}' }{ { S }_{ a } } \UNINEj{ { S }_{ 1 } }{ { S }_{0} }{ { S }_{ 1 }' }{ \half }{ { S }_{\rm p} }{ \half }{ S }{ S_{0}' }{ S' } \nonumber \\
      &\times \sum_{ { \eta }_{ b } { \eta }_{ a } } \sum_{ \substack{ { {\omega} }_{ b } { {\omega} }_{ a } \\ { \omega }_{0} { \omega }_{\rm p} } } \sum_{ \substack{ \rho_{0}' { \rho }_{ 0 } \\ \rho' { \rho }_{b} } } \sqrt{ \frac{ \text{dim}\, { \omega }_{0} }{ \text{dim}\, ({ \eta }_{ b }\, 0) } } \frac{ \text{dim}\, { \omega }_{\rm p} \, \text{dim}\, { {\omega} }_{ a } }{ \text{dim}\, {\omega}_{0}' \, \text{dim}\, (\eta ' \, 0) } \SuThreeUNINEFirst \SuThreeUNINESecond \nonumber \\
      & \times \sqrt{ 1 + { \delta }_{ { \eta }_{ a } \eta ' } } \sqrt{ 1 + { \delta }_{ { \eta }_{ a } \eta } } \sum_{ {\kappa}_{0}' S_{0}' } \SuThreeCGReduced{ \omega }{ \kappa L }{ {\omega}_{0} '}{ {\kappa }_{0}' S_{0}' }{ \omega' }{ \kappa' L' }{ \rho' } \frac{ { \Pi }_{ L' S' } }{ { \Pi }_{ S_{0}' } } { (-1) }^{ L + S_{0}' + S' + J } \WignerSIXj{ L }{ S_{0}' }{ L' }{ S' }{ J }{ S } \nonumber \\
      & \times \bra{ ({ \eta }_{ a } 0) (\eta ' 0) ; { {\omega} }_{ a } { S }_{ a } } | { \hat{V} }^{ {\omega}_{0} 'S_{0}' } | { \ket{ ({ \eta }_{ b } 0) (\eta 0) ; { {\omega} }_{ b } { S }_{ b } } }_{ \rho_{0}' } { \rho }_{ { \eta }_{ a } { \eta }_{ b } }^{ { \rho }_{ 0 }  { \omega }_{0} { S }_{0} } \left( {\nu}_{1}' {I}_{1}'; \nu_1 {I}_{1} \right),
      \label{directH}
  \end{align}
\end{widetext}
where $\omega_{\rm p}$, $\omega_{0}$, and $\omega_{0}'$ denote the SU(3) rank of the operator that transforms  the initial state to the final state of the projectile, target, and the $A+1$ system, respectively.
}
    \begin{figure}[th]
    \centering
        \centering
	\includegraphics[width=8.0cm]{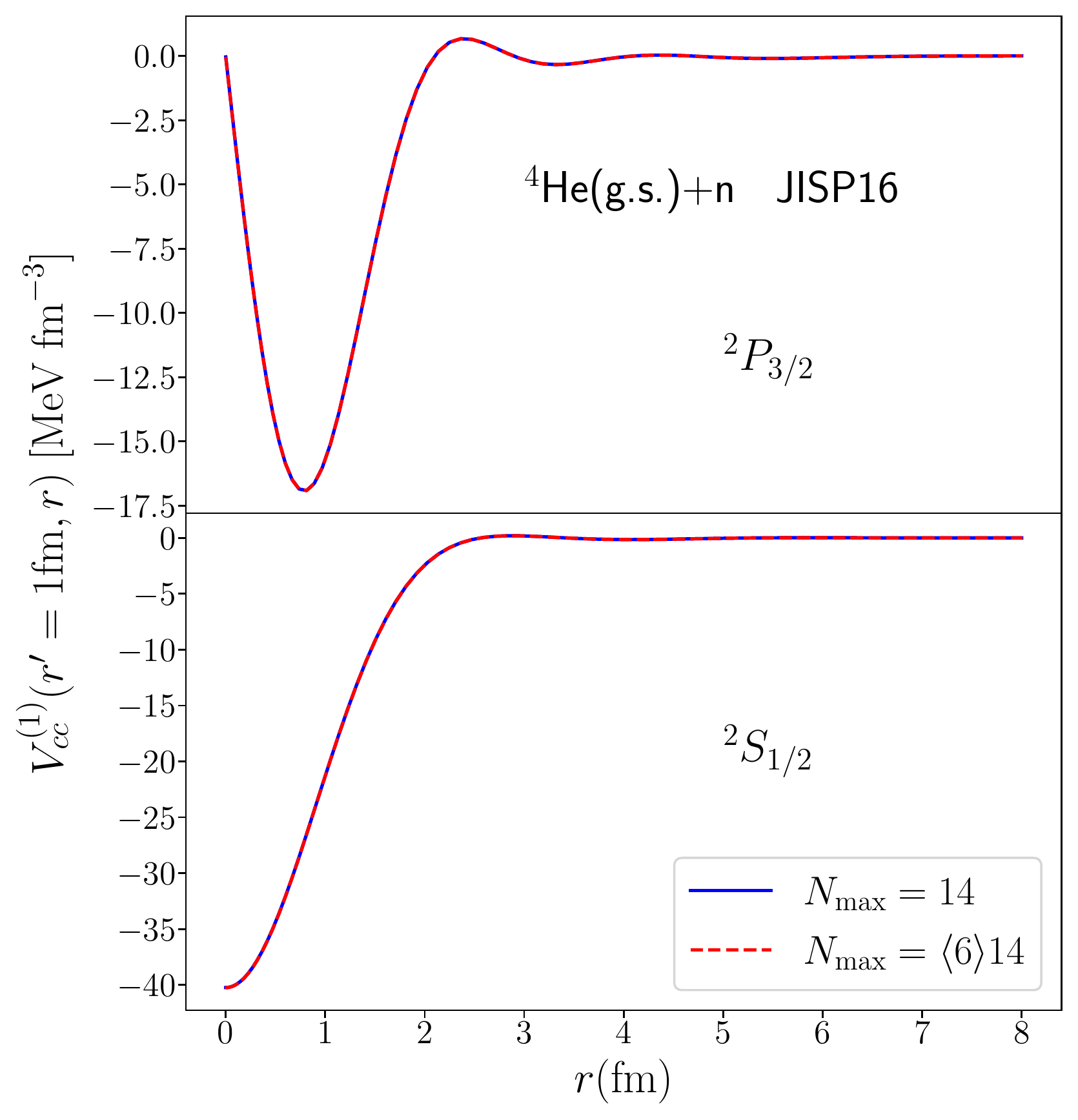} 
	\caption{Translationally invariant potential kernel of Eq. (\ref{directH}) for ${ {  }^{ 4 } }$He(${ 0 }^{ + }_{\rm g.s.}$) + n calculated with the JISP16 NN interaction, for ${ \hbar \Omega = 25 }$ MeV and $\eta_{\max}=10$, and using SA-NCSM ${ {  }^{ 4 } }$He wave functions in selected ($N_{\rm max}=\langle 6 \rangle 14$) and complete ($N_{\rm max}=14$) model spaces. 
	The selected space (dashed red) yields results that are indistinguishable from those in the complete space (solid blue).
	}
    \label{HKernel_4He_n}
\end{figure}

\section{Results and discussions}

It is important to validate the use of the SA basis in the SA-RGM, by comparing selected model spaces with the corresponding complete $N_{\rm max}$, to ensure that the selection does not remove configurations relevant for the reaction processes under consideration. For this, we study single-projectile scattering off the spherical $^4$He and  $^{16}$O nuclei, as well as for the deformed $^{20}$Ne nucleus. We present kernels that  use target ground state (g.s.) wavefunctions computed with the SA basis in a complete $N_{\rm max}$ model space (equivalent to NCSM/RGM calculations \cite{QuaglioniN09}) and we compare these to the results that use  wavefunctions calculated in a selected SA model space.
In general, SA selections are denoted as $\langle N_{\max}^{\rm C} \rangle N_{\max}$. For example, the ${ { N }_{ \text{max} } = \langle 6 \rangle 14 }$ model space includes the complete set of excitations up to ${ 6\hw }$ and selected excitations in the $8\hw$ - $14 \hw$ subspaces, following a prescription detailed in Ref. \cite{LauneyDSBD20}. This allows the mixing of all possible shapes within the complete subspaces, whereas the higher selected subspaces accommodate spatially expanded collective modes \cite{LauneyMD_ARNPS21}.
  \begin{figure}[th]
      \centering
          \centering
  	\includegraphics[width=1.\columnwidth]{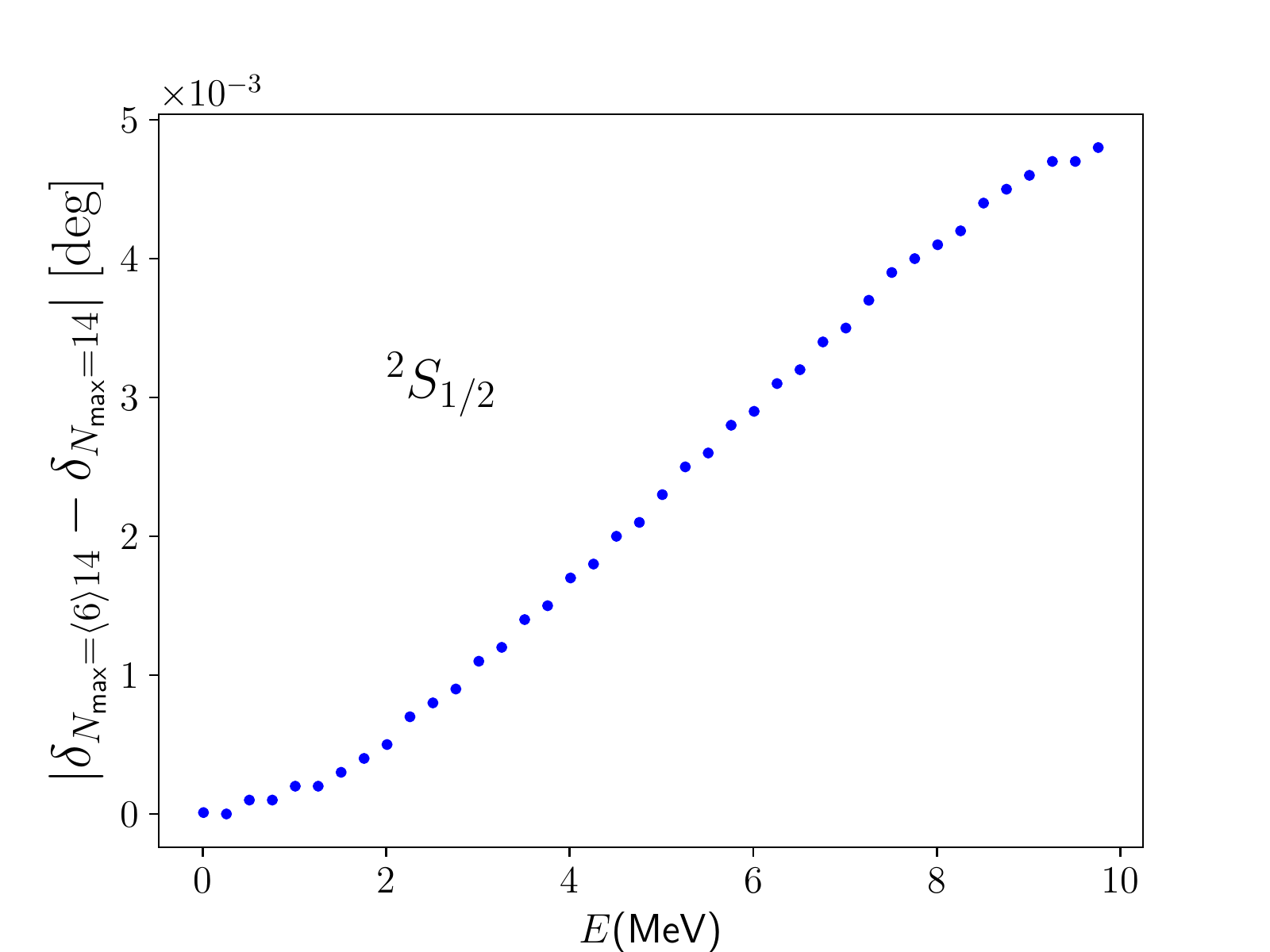} 
	\caption{Absolute value of the difference between the phase shifts ($\delta$) for the $^2S_{1/2}$ neutron scattering off ${^{4}}$He vs. the center-of-mass projectile energy, obtained in ${ {N}_{\text{max}} = \langle 6 \rangle 14 }$ and ${ {N}_{\text{max}} = 14 }$ model spaces.}
      \label{error_phase_shift_4He}
  \end{figure}
  \begin{figure*}[th]
    \centering
        \centering
	\includegraphics[width=8.0cm]{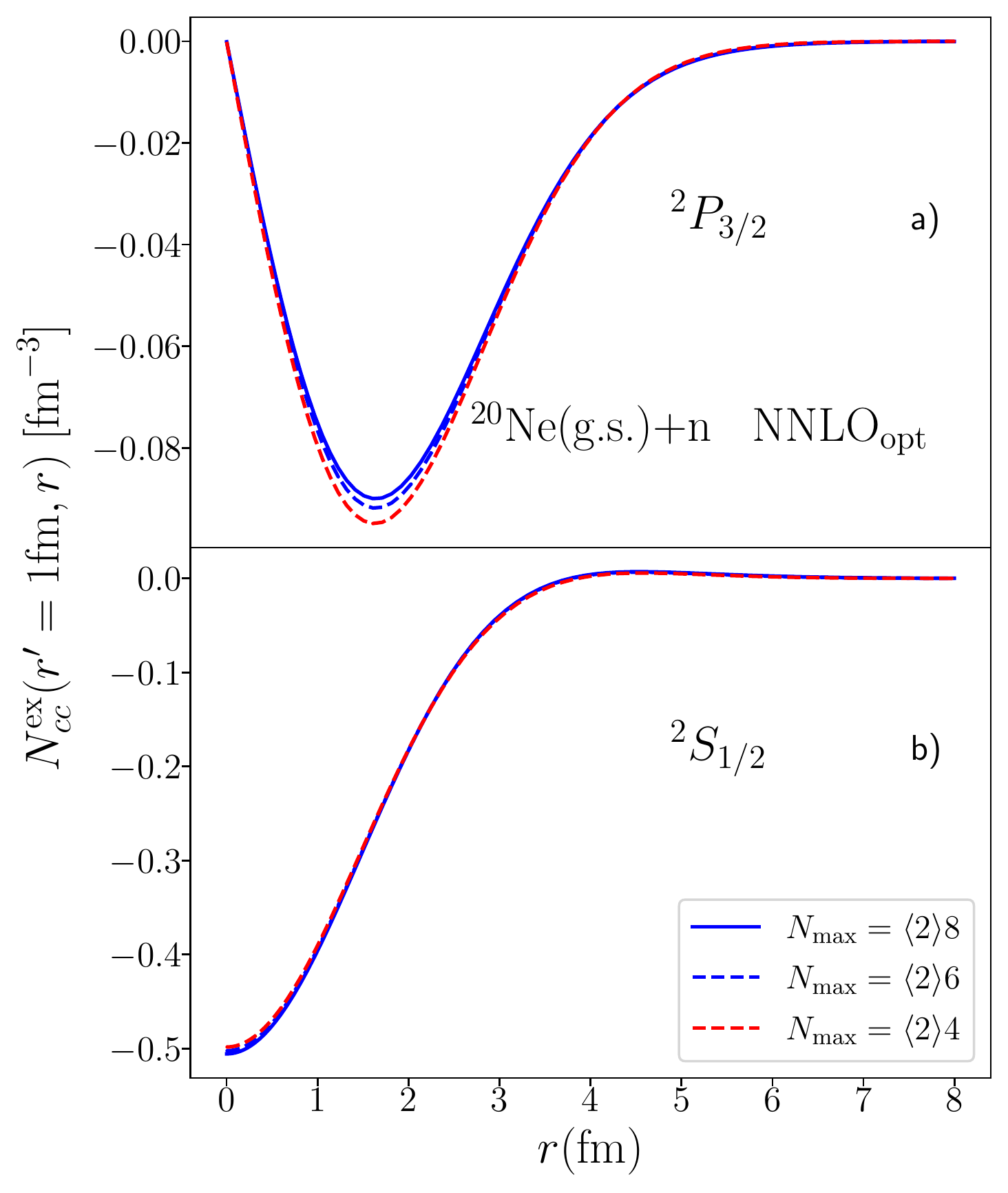} \includegraphics[width=8.0cm]{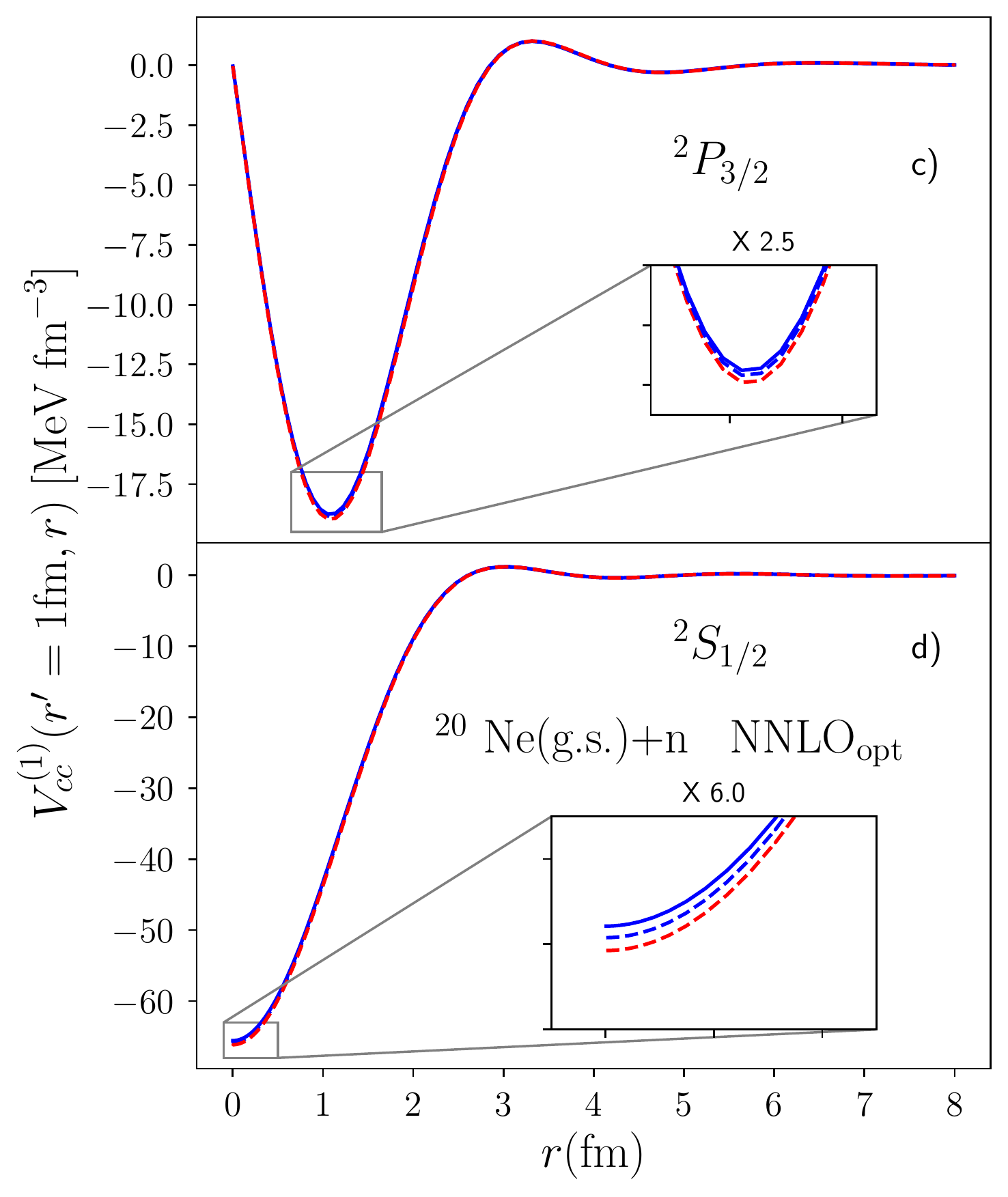} 
	\caption{
	Translationally invariant (a \& b) exchange norm kernel of Eq. (\ref{ExchangeMatrixSU3}) and (c \& d) potential kernel [Eqs. (\ref{dHKernel}) and (\ref{directH})] for ${ {  }^{ 20 } }$Ne(${ 0 }^{ + }_{\rm g.s.}$) + n calculated with the NNLO$_{\rm opt}$ NN interaction, for ${ \hbar \Omega = 15 }$ MeV and $\eta_{\max}=10$, and using SA-NCSM ${ {  }^{ 20 } }$Ne wave functions in selected $N_{\rm max}=\langle 2 \rangle 4$, $\langle 2 \rangle 6$, and $\langle 2 \rangle 8$ model spaces, for (a \& c) ${ ^2{ P }_{ 3/2 } }$ and (b \& d) ${ ^2{ S }_{ 1/2 } }$ partial waves. }
    \label{NKernel_20Ne_n}
\end{figure*}

\begin{figure}[th]
      \centering
          \centering
  	\includegraphics[width=1.\columnwidth]{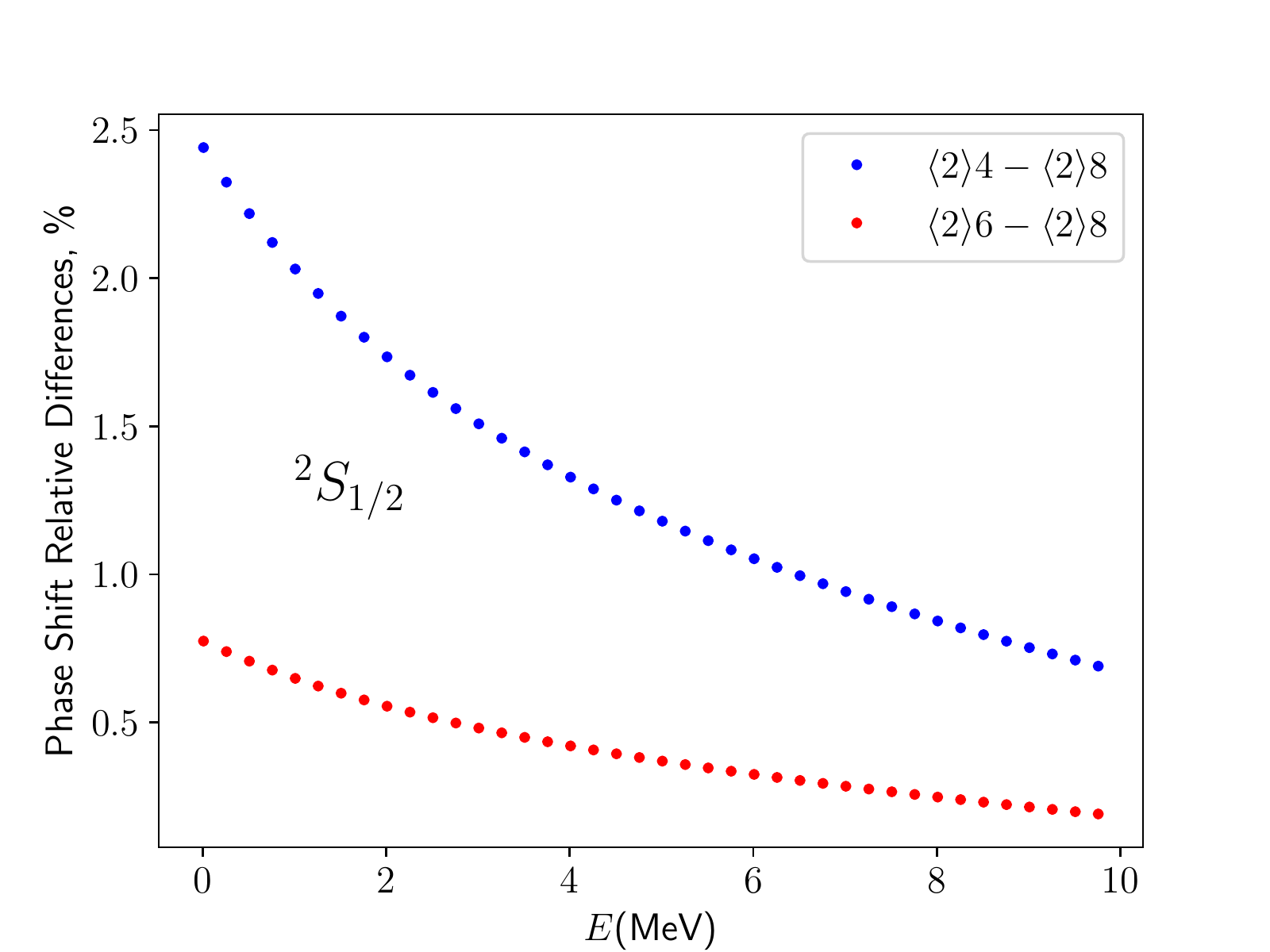} 
	\caption{
	Difference in phase shifts for the $^2S_{1/2}$ neutron scattering off ${ {}^{20} }$Ne from kernels calculated in ${ {N}_{\text{max}} = \langle 2 \rangle 4 }$ and ${ {N}_{\text{max}} = \langle 2 \rangle 6 }$ model spaces, relative to the phase shift from the largest model space ${ {N}_{\text{max}} = \langle 2 \rangle 8 }$. Results are shown as a function of the center-of-mass projectile energy.}
      \label{error_phase_shift_20Ne}
  \end{figure}

\subsection{Validation of the SA scheme}
\label{SAval}
{
  
  We study the SA efficacy for the potential kernel of Eq. (\ref{dHKernel}) for ${ {  }^{ 4 } }$He(${ { 0 }^{ + }_{\rm g.s.} }$) + n (Fig. \ref{HKernel_4He_n}), for which calculations in the complete space (no SA selection) are available up to $N_{\rm max}=18/19$ with other interactions ($N_{\rm max}=18$ denotes the model space for the target) \cite{QuaglioniN09}. For two NN interactions, Ref. \cite{QuaglioniN09} has shown that the $N_{\rm max}=14/15$ results are sufficient to achieve converged phase shifts for the ${ {  }^{ 4 } }$He(${ { 0 }^{ + }_{\rm g.s.} }$) + n $^2{ { S }_{ 1/2 } }$ and $^2{ { P }_{ 3/2 } }$ channels. In the present study,
  we use the ${ { N }_{ \text{max} } = 14 }$ complete model space for the target, and we compare to the ${ { N }_{ \text{max} } = \langle 6 \rangle 14 }$ model space.  The ${ {  }^{ 4 } }$He wavefunctions in these model spaces, calculated with the JISP16  NN interactions \cite{ShirokovMZVW07}, have been shown to converge for the binding energy and the g.s. root-mean-square (rms) matter radius, as well as to yield various electromagnetic sum rules \cite{baker_2020} that agree with those calculated in the hyperspherical harmonics  approach \cite{bacca:2013_prl}.

  We explore the potential kernel of Eq. (\ref{dHKernel}) for ${ { P }_{ 3/2 } }$ as a function of the distance between the clusters, which is used to describe the  ${ { \frac{ 3 }{ 2 } }^{ - } }$  resonant g.s. in ${ {  }^{ 5 } }$He (Fig. \ref{HKernel_4He_n}a), as well as for  ${ { S }_{ 1/2 } }$ for a description of the ${ { \frac{ 1 }{ 2 } }^{ + } }$ scattering states of ${ {  }^{ 5 } }$He (Fig. \ref{HKernel_4He_n}b). 
  We find that the SA space yields results that are indistinguishable from those in the complete space. While Fig. \ref{HKernel_4He_n} shows the comparison only for $r'=1$ fm, the results remain indistinguishable for any $r'$. In addition, the norm kernels exhibit the same behavior, namely, the outcomes for the SA and complete model spaces coincide.
  These results demonstrate that the SA wavefunctions account for the relevant correlations necessary to describe the norm and the direct component of the non-local potentials that govern the  resonant ground state and low-energy scattering states in ${ {  }^{ 5 } }$He. Because the kernels are used as the input for calculating phase shifts, the findings show that the SA model spaces are sufficient to reproduce the corresponding ${ { S }_{ 1/2 } }$ and ${ { P }_{ 3/2 } }$ phase shifts calculated in the $N_{\rm max}$ complete model spaces (see Fig. \ref{error_phase_shift_4He} for the comparison between selected and complete model spaces for the $S$ wave). We emphasize that this comparison focuses on the effect of SA model spaces benchmarked against the corresponding complete model spaces, not on reproducing experimental phase shifts with all RGM kernels.

\subsection{Application to intermediate-mass nuclei}
\label{SAappl}
   To illustrate the capability of the SA-RGM, we present the first \textit{ab initio} calculations of RGM norm and leading-order potential kernels in the intermediate mass region, namely, for neutron scattering off ${ {  }^{ 20 } }$Ne(${ 0 }^{ + }_{\rm g.s.}$).
  The SA ${ {  }^{ 20 } }$Ne wave functions are calculated using the NNLO$_{\rm opt}$ NN interaction \cite{Ekstrom13} and have been shown to reproduce observables, such as excitation energies and B(E2) strengths \cite{DytrychLDRWRBB20}. The NNLO$_{\rm opt}$ is used without 3N forces, which have been shown to contribute minimally to the 3- and 4-nucleon binding energies \cite{Ekstrom13}. Furthermore, the NNLO$_{\rm opt}$ NN potential has been found to reproduce various observables, including the $^4$He electric dipole polarizability \cite{BakerLBND20}; the challenging analyzing power for elastic proton scattering on $^4$He, $^{12}$C, and $^{16}$O \cite{BurrowsEWLMNP19}; along with  B(E2) transition strengths  for $^{21}$Mg and $^{21}$F  \cite{Ruotsalainen19}.

  As expected,  the exchange norm kernel for ${ {  }^{ 20 } }$Ne(${ 0 }^{ + }_{\rm g.s.})+{\rm n}$
  manifests itself at short distances and vanishes at long distances (see Fig. \ref{NKernel_20Ne_n}a \& b, for the case of $r'=1$ fm).
  This reflects the short-range nature of the Pauli exclusion principle.
  We find that the change in the model space size from $N_{\rm max}=6$ to $N_{\rm max}=8$ has only a small effect on the exchange norm of the ${ { P }_{ 3/2 } }$ partial wave  (Fig. \ref{NKernel_20Ne_n}a) and ${ { S }_{ 1/2 } }$ partial wave  (Fig. \ref{NKernel_20Ne_n}b). The largest deviations are observed at short distances, where the kernels have the largest magnitude. As the model space increases, the kernels start to converge, and the exchange kernel maximum slightly increases in magnitude for ${ { S }_{ 1/2 } }$, whereas it slightly decreases for ${ { P }_{ 3/2 } }$.
  Note that even though the deviation seems larger for ${ { P }_{ 3/2 } }$, the magnitude of the ${ { P }_{ 3/2 } }$ exchange kernel maximum is smaller by a factor of 3.5 than that of ${ { S }_{ 1/2 } }$. 
  Hence, these outcome indicates that the selection of dominant SU(3) components  at $N_{\rm max}=8$ (see also Fig. 3 in Ref.~\cite{DytrychLDRWRBB20}) is sufficient to incorporate the relevant correlations needed to describe the short-range Pauli effect.

  The potential kernels of Eq. (\ref{dHKernel}) for ${ {  }^{ 20 } }$Ne(${ 0 }^{ + }_{\rm g.s.})+{\rm n}$, calculated with the NNLO$_{\rm opt}$ NN, are also studied with increasing model space sizes (see Fig. \ref{NKernel_20Ne_n}c \& d, for $r'=1$ fm).
  Similarly to the exchange norm kernel, the increase in the model space from $N_{\rm max}=6$ to $N_{\rm max}=8$ has a much smaller effect on this potential kernel when compared to the increase from $N_{\rm max}=4$ to $N_{\rm max}=6$, for both  ${ { S }_{ 1/2 } }$ and ${ { P }_{ 1/2 } }$ partial waves, suggesting converging results. 
   When compared to the potentials for ${ {  }^{ 4 } }$He + n of Fig. \ref{NKernel_20Ne_n}, the ${ {  }^{ 20 } }$Ne(${ 0 }^{ + }_{\rm g.s.})+{\rm n}$ case shows a slightly larger deviation around the kernel maximum when varying the model space.
  This effect might be a result of the open-shell structure of the ground-state wave function of ${ {  }^{ 20 } }$Ne compared to that of  ${ {  }^{ 4 } }$He.
  In addition, the small changes in these kernels result in only very little deviations in the $^2S_{1/2}$ phase shifts for the low-energy neutron scattering off ${ {  }^{ 20 } }$Ne(${ 0 }^{ + }_{\rm g.s.})$, with a relative difference of the order of $1$-$2\%$ compared to the largest model space used (Fig. \ref{error_phase_shift_20Ne}).

  As another illustrative example, we present the potential kernel of Eq. (\ref{dHKernel}) for ${ {  }^{ 16 } }$O(${ 0 }^{ + }_{\rm g.s.})+{\rm n}$ (Fig. \ref{HKernel_16O_n}), which is feasible for no-core shell-model calculations with the importance truncation using other interactions \cite{PhysRevC.82.034609}. In our study, we use the NNLO$_{\rm sat}$ \cite{PhysRevC.91.051301}, for which the three-nucleon (3N) forces are included in the SA-NCSM as averages \cite{LauneyMD_ARNPS21}. Namely, in these calculations, the 3N forces are included as a mass-dependent monopole interaction \cite{LauneyDD12}, which  has an effect on binding energies. or the $^{16}$O ground-state energy,  
the  7-shell 3N contribution is 20.46 MeV, resulting in $-127.97$ MeV total energy for $N_{\rm max}=8$ and \hw=16 MeV, which agrees with the experimental value of $-127.62$ MeV.
  In this case, we compare calculations within a selected model space ${ { N }_{ \text{max} } = \langle 0 \rangle 8 }$ to those in the complete ${ { N }_{ \text{max}} = 6   }$ model space.
  The results of the two model spaces are practically indistinguishable, despite the largely reduced SA model space used here and the addition of SU(3) dominant configurations in the 8\hw~subspace. For $^{16}$O, this outcome could be understood by the fact that $\sim 80$\% of the ground state is composed of a spherical shape and low $N_{\rm max}$ model spaces are able to account for its vibrations.

  \begin{figure}[th]
    \centering
        \centering
	\includegraphics[width=8.0cm]{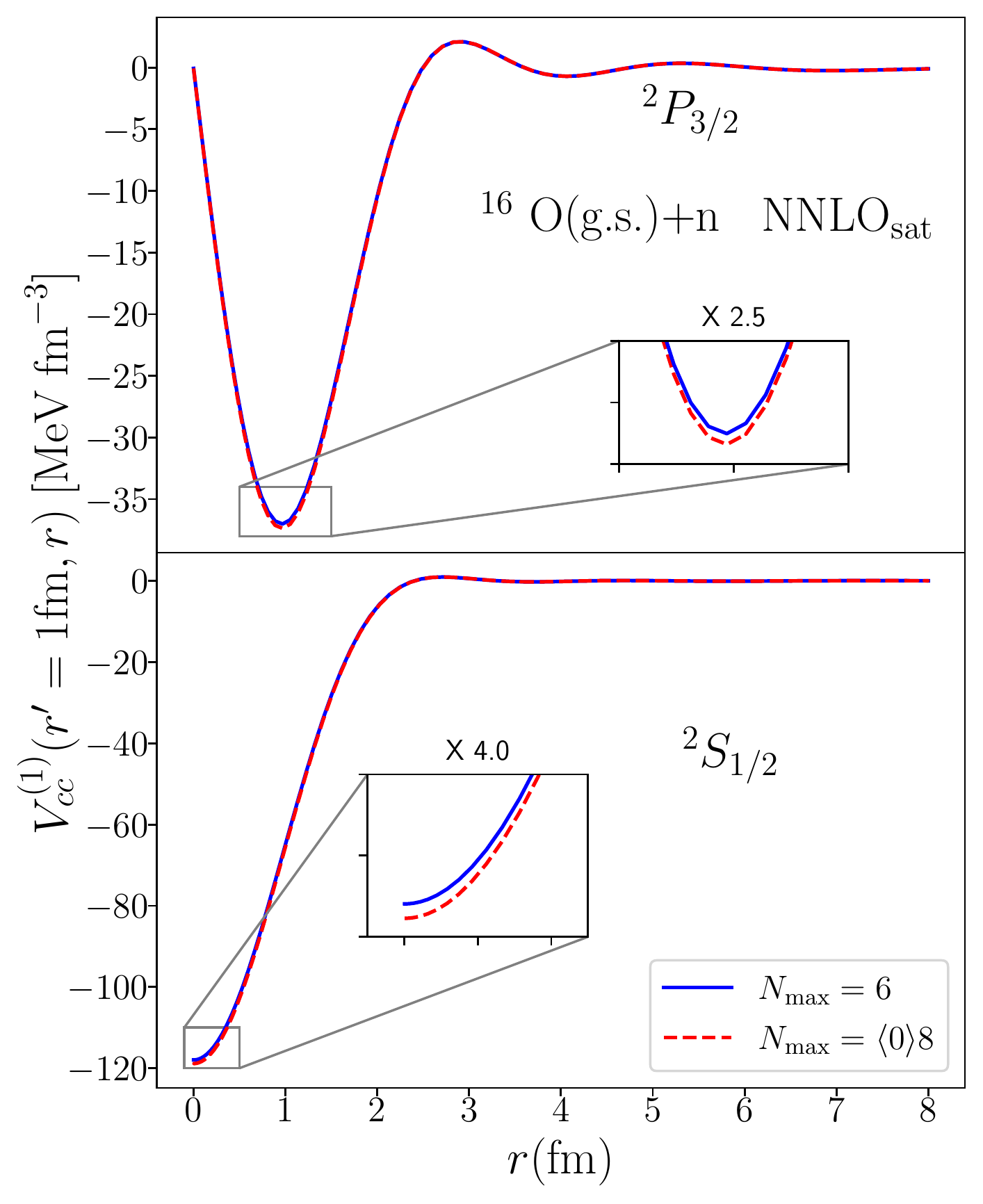} 
	\caption{Translationally invariant potential kernel [Eqs. (\ref{dHKernel}] and (\ref{directH})) for ${ {  }^{ 16 } }$O + n calculated with the NNLO$_{\rm sat}$ NN+3N interaction, for ${ \hbar \Omega = 16 }$ MeV and $\eta_{\max}=10$, and using SA-NCSM ${ {  }^{ 16 } }$O wave functions in selected ($N_{\rm max}=\langle 0 \rangle 8$) and complete ($N_{\rm max}=6$) model spaces. }
    \label{HKernel_16O_n}
\end{figure}

}

\subsection{Efficacy and scalability of the SA scheme}
\label{dimension_analysis}
{

  In this section we explore the scalability of the SA-RGM calculations  with increasing model space sizes and particle number. 
  The SA-RGM channel basis (\ref{SU3RGMstates}) is used to compute the kernels of Eqs.~(\ref{ExchangeMatrixSU3}) and (\ref{directH}).
  These channels are constructed from the unique ${ \left\{ \nu_{ 1 } \right\} }$ quantum numbers of the target state, resulting in a manageable number of  SA-RGM basis states that scale polynomially with $N_{\rm max}$, as shown in Fig. \ref{dim_D_Phi_A5_1I2+} for several nuclear systems.
 
   For example, for proton- or neutron-nucleus interaction for N+$^{20}$Ne (${ 0 }^{ + }_{\rm g.s.}$), there are only about $10^3$-$10^4$ SA-RGM basis  states for 7 to 13 shells, and only about $10^5-10^6$ for $^{23}$Mg when more target states are used (with channels for ${ 3/2_{\rm g.s.} ^{ + } }, { { 5/2 }^{ + } }$, ${ { 7/2 }^{ + } }$), which is still manageable (Fig. \ref{dim_D_Phi_A5_1I2+}). 
 Interestingly, the number of unique deformed configurations for heavier targets, such as Ne and Mg, decrease in larger model spaces, as dominant shapes are allowed to develop, thereby reducing shape mixing.
  As a consequence, in such cases the SA-RGM basis can become smaller when increasing ${ { N }_{ \text{max} } }$.

Furthermore, there is a large reduction in the number of SU(3) basis states needed for the target wave functions, as one eliminates negligible contributions identified in the target eigenfunctions. 
  Namely, for the illustrative example of the ${ {  }^{ 23 } }$Mg target (Fig. \ref{dim_D_Phi_A5_1I2+}), we show the number of the SA-RGM channels after retaining basis states that contribute with a probability amplitude $({ C }_{ { \mathfrak{b} }_1 }^{ { \omega }_{ 1 } { \kappa }_{ 1 } L_1{ S }_{ 1 } } )^2$ [see Eq. (\ref{SU3wf})] greater than a certain value $\varepsilon$. We find that the number of the ${^{23}}$Mg+N SA-RGM states continues to scale polynomially with increasing ${ { N }_{ \text{max} } }$ for each $\varepsilon$ and largely decreases for higher $\varepsilon$ reduction cutoffs.
  We note that the $\varepsilon=10^{-6}$ cutoff uses the ${ {  }^{ 23 } }$Mg basis states with a probability greater that ${ { 10 }^{ -6 } }$ and results in no reduction. In the SA-RGM calculations, the $\varepsilon$ cutoff for the SA selection is decreased until convergence of results is achieved.
  
  We note that an important step for computing the kernels from the many-body wavefunctions is the calculation of the ${\rho}_{ \eta  \eta' }^{  { \rho }_{ 0} { \omega }_{0} { S }_{0} }$ operator  of Eq. (\ref{O_objects}).  Its calculation can be compared to the one-body density matrix elements, namely, they need to be calculated only once for a given set of target wavefunctions, and, as mentioned above, can utilize an efficient algorithm that exploits SU(3) SA subspaces and the factorization of spatial and spin degrees of freedom.
  As for the kernels, these calculations are also facilitated by the large reduction in the number of SU(3) basis states needed to describe the target wave functions, as compared to the complete $N_{\rm max}$ model space. These same reductions are observed for two-body densities that will be needed for the particle-rank two potential kernels.
    \begin{figure}[th]
      \centering
          \centering
  	\includegraphics[width=1.\columnwidth]{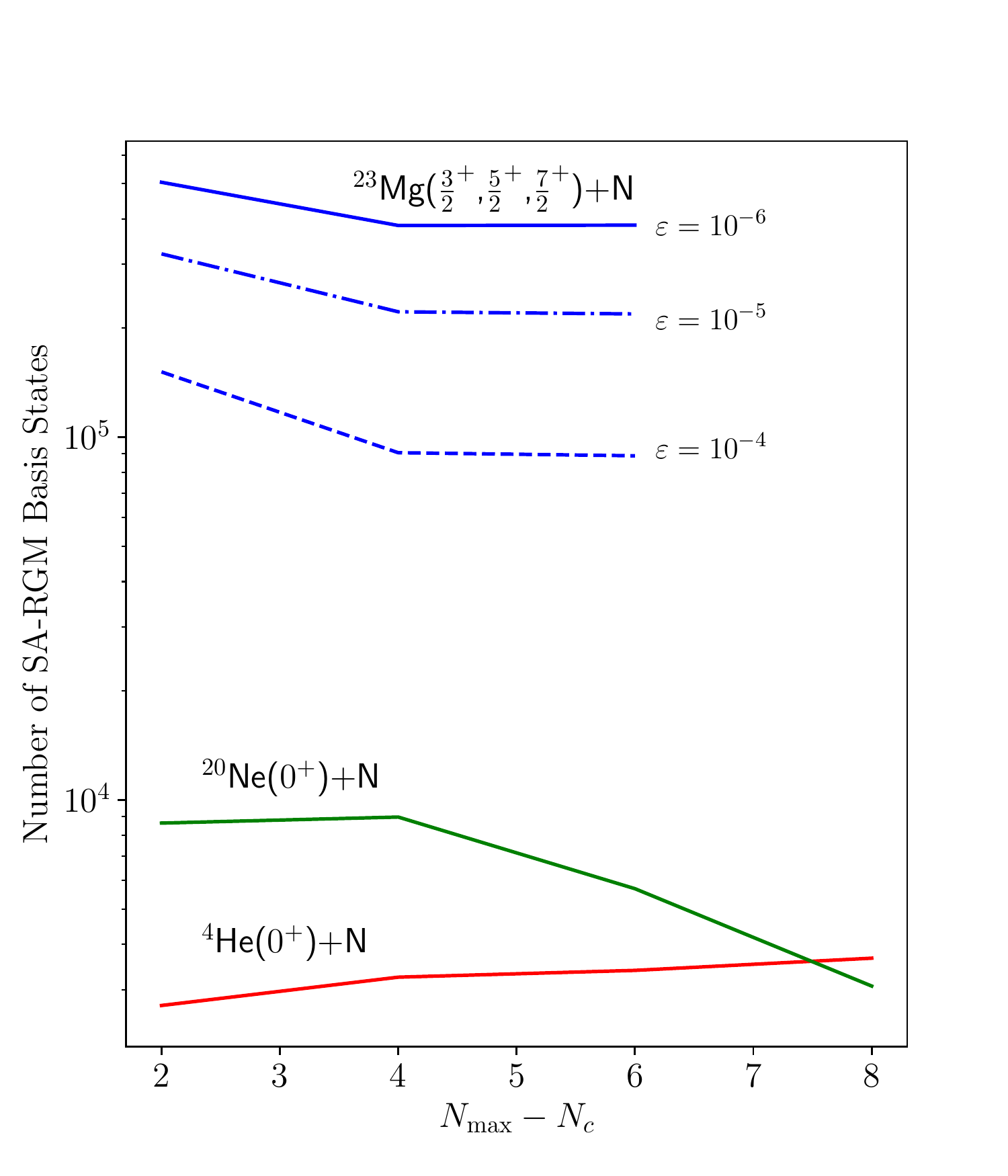} 
	\caption{
	Number of SA-RGM basis states as a function of the model space size ${\langle N_{\rm max}^{\rm C} \rangle  { N }_{ \text{max} } }$ of the target. The target eigenfunctions are shown as complete (solid curves or $\varepsilon=10^{-6}$), or reduced to the SU(3) basis states with probability amplitudes greater than $\varepsilon$ cutoff. We use $N_{\rm max}^{\rm C}=6$ for ${ {  }^{ 4 } }$He, and $N_{\rm max}^{\rm C}=2$ for ${ {  }^{ 20 } }$Ne and ${ {  }^{ 23 } }$Mg, as well as $\eta_{\rm max}=15$.  The SA-NCSM calculations for ${ {  }^{ 4 } \text{He}}$ (${ {  }^{ 20 } \text{Ne}}$ and ${ {  }^{ 23 } \text{Mg} }$) use the JISP16 (N2LOopt) NN interaction and ${ \hbar \Omega = 25 }$ MeV (${ \hbar \Omega = 25 }$ and ${ \hbar \Omega = 15 }$, respectively).
	}
      \label{dim_D_Phi_A5_1I2+}
  \end{figure}

}

\section{Conclusions}
\label{concl}
In this paper, we have studied the efficacy of the new \textit{ab initio} SA-RGM approach that combines the SA-NCSM and RGM frameworks. We have discussed nucleon-nucleus interactions and the use of the SA framework for $^{4}$He and $^{16}$O targets, as well as the intermediate-mass $^{20}$Ne and $^{23}$Mg targets feasible in the SA-NCSM. We have shown that the SU(3) selection of the model space has almost negligible effect on the SA-RGM norm and particle-rank one potential kernels that provide the input to calculations of phase shifts and cross sections. The results demonstrate that the nonnegligible components that are included in the calculations account for the correlations needed to describe the single nucleon scattering process in this mass region. 

In addition, we have studied the scalability of the SA-RGM approach,  showing its computational advantages that stem from the largely reduced number of SU(3) basis states needed to describe the target, as well as the manageable number of the SA-RGM basis states for the target+N system that scale polynomially with the increase in the model space size. The demonstrated efficacy of the SA basis and its scalability with particle numbers and model space dimensions opens the way to \textit{ab initio} calculations up through the medium-mass region of nucleon-nucleus interactions that enter nucleon scattering and  nucleon capture reactions.

\begin{acknowledgments}
We acknowledge useful discussions with Petr Navratil, as well as  Linda Hlophe for calculating the $N_{\rm max}=4$ NCSM/RGM kernel matrix elements for model benchmarks. This work was supported in part by the U.S. National Science Foundation  (PHY-1913728), SURA, the Czech Science Foundation (16-16772S), and the U.S. Department of Energy, Office of Science, Office of Nuclear Physics, under Work Proposals No. SCW0498 and No. SC0019521. A portion of this work was performed under the auspices of the U.S. Department of Energy by Lawrence Livermore National Laboratory under Contract No. DE-AC52-07NA27344 with support from LDRD Project No. 19-ERD-017. It benefited from high performance computational resources provided by LSU (www.hpc.lsu.edu),  the National Energy Research Scientific Computing Center (NERSC), a U.S. Department of Energy Office of Science User Facility operated under Contract No. DE-AC02-05CH11231, as well as the Frontera computing project at the Texas Advanced Computing Center,  made possible by National Science Foundation award OAC-1818253.

\end{acknowledgments}

\bibliographystyle{apsrev4-1}
\bibliography{refs}

\begin{thebibliography}{78}%
\makeatletter
\providecommand \@ifxundefined [1]{%
 \@ifx{#1\undefined}
}%
\providecommand \@ifnum [1]{%
 \ifnum #1\expandafter \@firstoftwo
 \else \expandafter \@secondoftwo
 \fi
}%
\providecommand \@ifx [1]{%
 \ifx #1\expandafter \@firstoftwo
 \else \expandafter \@secondoftwo
 \fi
}%
\providecommand \natexlab [1]{#1}%
\providecommand \enquote  [1]{``#1''}%
\providecommand \bibnamefont  [1]{#1}%
\providecommand \bibfnamefont [1]{#1}%
\providecommand \citenamefont [1]{#1}%
\providecommand \href@noop [0]{\@secondoftwo}%
\providecommand \href [0]{\begingroup \@sanitize@url \@href}%
\providecommand \@href[1]{\@@startlink{#1}\@@href}%
\providecommand \@@href[1]{\endgroup#1\@@endlink}%
\providecommand \@sanitize@url [0]{\catcode `\\12\catcode `\$12\catcode
  `\&12\catcode `\#12\catcode `\^12\catcode `\_12\catcode `\%12\relax}%
\providecommand \@@startlink[1]{}%
\providecommand \@@endlink[0]{}%
\providecommand \url  [0]{\begingroup\@sanitize@url \@url }%
\providecommand \@url [1]{\endgroup\@href {#1}{\urlprefix }}%
\providecommand \urlprefix  [0]{URL }%
\providecommand \Eprint [0]{\href }%
\providecommand \doibase [0]{http://dx.doi.org/}%
\providecommand \selectlanguage [0]{\@gobble}%
\providecommand \bibinfo  [0]{\@secondoftwo}%
\providecommand \bibfield  [0]{\@secondoftwo}%
\providecommand \translation [1]{[#1]}%
\providecommand \BibitemOpen [0]{}%
\providecommand \bibitemStop [0]{}%
\providecommand \bibitemNoStop [0]{.\EOS\space}%
\providecommand \EOS [0]{\spacefactor3000\relax}%
\providecommand \BibitemShut  [1]{\csname bibitem#1\endcsname}%
\let\auto@bib@innerbib\@empty
%</preamble>
\bibitem [{\citenamefont {Launey}\ \emph {et~al.}(2016)\citenamefont {Launey},
  \citenamefont {Dytrych},\ and\ \citenamefont {Draayer}}]{LauneyDD16}%
  \BibitemOpen
  \bibfield  {author} {\bibinfo {author} {\bibfnamefont {K.~D.}\ \bibnamefont
  {Launey}}, \bibinfo {author} {\bibfnamefont {T.}~\bibnamefont {Dytrych}}, \
  and\ \bibinfo {author} {\bibfnamefont {J.~P.}\ \bibnamefont {Draayer}},\
  }\href {\doibase 10.1016/j.ppnp.2016.02.001} {\bibfield  {journal} {\bibinfo
  {journal} {Prog. Part. Nucl. Phys.}\ }\textbf {\bibinfo {volume} {89}},\
  \bibinfo {pages} {101 (review)} (\bibinfo {year} {2016})}\BibitemShut
  {NoStop}%
\bibitem [{\citenamefont {Dytrych}\ \emph {et~al.}(2020)\citenamefont
  {Dytrych}, \citenamefont {Launey}, \citenamefont {Draayer}, \citenamefont
  {Rowe}, \citenamefont {Wood}, \citenamefont {Rosensteel}, \citenamefont
  {Bahri}, \citenamefont {Langr},\ and\ \citenamefont
  {Baker}}]{DytrychLDRWRBB20}%
  \BibitemOpen
  \bibfield  {author} {\bibinfo {author} {\bibfnamefont {T.}~\bibnamefont
  {Dytrych}}, \bibinfo {author} {\bibfnamefont {K.~D.}\ \bibnamefont {Launey}},
  \bibinfo {author} {\bibfnamefont {J.~P.}\ \bibnamefont {Draayer}}, \bibinfo
  {author} {\bibfnamefont {D.~J.}\ \bibnamefont {Rowe}}, \bibinfo {author}
  {\bibfnamefont {J.~L.}\ \bibnamefont {Wood}}, \bibinfo {author}
  {\bibfnamefont {G.}~\bibnamefont {Rosensteel}}, \bibinfo {author}
  {\bibfnamefont {C.}~\bibnamefont {Bahri}}, \bibinfo {author} {\bibfnamefont
  {D.}~\bibnamefont {Langr}}, \ and\ \bibinfo {author} {\bibfnamefont {R.~B.}\
  \bibnamefont {Baker}},\ }\href {\doibase 10.1103/PhysRevLett.124.042501}
  {\bibfield  {journal} {\bibinfo  {journal} {Phys. Rev. Lett.}\ }\textbf
  {\bibinfo {volume} {124}},\ \bibinfo {pages} {042501} (\bibinfo {year}
  {2020})}\BibitemShut {NoStop}%
\bibitem [{\citenamefont {Dytrych}\ \emph {et~al.}(2015)\citenamefont
  {Dytrych}, \citenamefont {Hayes}, \citenamefont {Launey}, \citenamefont
  {Draayer}, \citenamefont {Maris}, \citenamefont {Vary}, \citenamefont
  {Langr},\ and\ \citenamefont {Oberhuber}}]{DytrychHLDMVLO14}%
  \BibitemOpen
  \bibfield  {author} {\bibinfo {author} {\bibfnamefont {T.}~\bibnamefont
  {Dytrych}}, \bibinfo {author} {\bibfnamefont {A.~C.}\ \bibnamefont {Hayes}},
  \bibinfo {author} {\bibfnamefont {K.~D.}\ \bibnamefont {Launey}}, \bibinfo
  {author} {\bibfnamefont {J.~P.}\ \bibnamefont {Draayer}}, \bibinfo {author}
  {\bibfnamefont {P.}~\bibnamefont {Maris}}, \bibinfo {author} {\bibfnamefont
  {J.~P.}\ \bibnamefont {Vary}}, \bibinfo {author} {\bibfnamefont
  {D.}~\bibnamefont {Langr}}, \ and\ \bibinfo {author} {\bibfnamefont
  {T.}~\bibnamefont {Oberhuber}},\ }\href {\doibase 10.1103/PhysRevC.91.024326}
  {\bibfield  {journal} {\bibinfo  {journal} {Phys. Rev. C}\ }\textbf {\bibinfo
  {volume} {91}},\ \bibinfo {pages} {024326} (\bibinfo {year}
  {2015})}\BibitemShut {NoStop}%
\bibitem [{\citenamefont {Baker}\ \emph
  {et~al.}(2020{\natexlab{a}})\citenamefont {Baker}, \citenamefont {Launey},
  \citenamefont {Bacca}, \citenamefont {Dinur},\ and\ \citenamefont
  {Dytrych}}]{BakerLBND20}%
  \BibitemOpen
  \bibfield  {author} {\bibinfo {author} {\bibfnamefont {R.~B.}\ \bibnamefont
  {Baker}}, \bibinfo {author} {\bibfnamefont {K.~D.}\ \bibnamefont {Launey}},
  \bibinfo {author} {\bibfnamefont {S.}~\bibnamefont {Bacca}}, \bibinfo
  {author} {\bibfnamefont {N.~N.}\ \bibnamefont {Dinur}}, \ and\ \bibinfo
  {author} {\bibfnamefont {T.}~\bibnamefont {Dytrych}},\ }\href {\doibase
  10.1103/PhysRevC.102.014320} {\bibfield  {journal} {\bibinfo  {journal}
  {Phys. Rev. C}\ }\textbf {\bibinfo {volume} {102}},\ \bibinfo {pages}
  {014320} (\bibinfo {year} {2020}{\natexlab{a}})}\BibitemShut {NoStop}%
\bibitem [{\citenamefont {Ruotsalainen}\ \emph {et~al.}(2019)\citenamefont
  {Ruotsalainen}, \citenamefont {Henderson}, \citenamefont {Hackman},
  \citenamefont {Sargsyan}, \citenamefont {Launey}, \citenamefont {Saxena},
  \citenamefont {Srivastava}, \citenamefont {Stroberg}, \citenamefont {Grahn},
  \citenamefont {Pakarinen}, \citenamefont {Ball}, \citenamefont {Julin},
  \citenamefont {Greenlees}, \citenamefont {Smallcombe}, \citenamefont
  {Andreoiu}, \citenamefont {Bernier}, \citenamefont {Bowry}, \citenamefont
  {Buckner}, \citenamefont {Caballero-Folch}, \citenamefont {Chester},
  \citenamefont {Cruz}, \citenamefont {Evitts}, \citenamefont {Frederick},
  \citenamefont {Garnsworthy}, \citenamefont {Holl}, \citenamefont {Kurkjian},
  \citenamefont {Kisliuk}, \citenamefont {Leach}, \citenamefont {McGee},
  \citenamefont {Measures}, \citenamefont {M\"ucher}, \citenamefont {Park},
  \citenamefont {Sarazin}, \citenamefont {Smith}, \citenamefont {Southall},
  \citenamefont {Starosta}, \citenamefont {Svensson}, \citenamefont {Whitmore},
  \citenamefont {Williams},\ and\ \citenamefont {Wu}}]{Ruotsalainen19}%
  \BibitemOpen
  \bibfield  {author} {\bibinfo {author} {\bibfnamefont {P.}~\bibnamefont
  {Ruotsalainen}}, \bibinfo {author} {\bibfnamefont {J.}~\bibnamefont
  {Henderson}}, \bibinfo {author} {\bibfnamefont {G.}~\bibnamefont {Hackman}},
  \bibinfo {author} {\bibfnamefont {G.~H.}\ \bibnamefont {Sargsyan}}, \bibinfo
  {author} {\bibfnamefont {K.~D.}\ \bibnamefont {Launey}}, \bibinfo {author}
  {\bibfnamefont {A.}~\bibnamefont {Saxena}}, \bibinfo {author} {\bibfnamefont
  {P.~C.}\ \bibnamefont {Srivastava}}, \bibinfo {author} {\bibfnamefont
  {S.~R.}\ \bibnamefont {Stroberg}}, \bibinfo {author} {\bibfnamefont
  {T.}~\bibnamefont {Grahn}}, \bibinfo {author} {\bibfnamefont
  {J.}~\bibnamefont {Pakarinen}}, \bibinfo {author} {\bibfnamefont {G.~C.}\
  \bibnamefont {Ball}}, \bibinfo {author} {\bibfnamefont {R.}~\bibnamefont
  {Julin}}, \bibinfo {author} {\bibfnamefont {P.~T.}\ \bibnamefont
  {Greenlees}}, \bibinfo {author} {\bibfnamefont {J.}~\bibnamefont
  {Smallcombe}}, \bibinfo {author} {\bibfnamefont {C.}~\bibnamefont
  {Andreoiu}}, \bibinfo {author} {\bibfnamefont {N.}~\bibnamefont {Bernier}},
  \bibinfo {author} {\bibfnamefont {M.}~\bibnamefont {Bowry}}, \bibinfo
  {author} {\bibfnamefont {M.}~\bibnamefont {Buckner}}, \bibinfo {author}
  {\bibfnamefont {R.}~\bibnamefont {Caballero-Folch}}, \bibinfo {author}
  {\bibfnamefont {A.}~\bibnamefont {Chester}}, \bibinfo {author} {\bibfnamefont
  {S.}~\bibnamefont {Cruz}}, \bibinfo {author} {\bibfnamefont {L.~J.}\
  \bibnamefont {Evitts}}, \bibinfo {author} {\bibfnamefont {R.}~\bibnamefont
  {Frederick}}, \bibinfo {author} {\bibfnamefont {A.~B.}\ \bibnamefont
  {Garnsworthy}}, \bibinfo {author} {\bibfnamefont {M.}~\bibnamefont {Holl}},
  \bibinfo {author} {\bibfnamefont {A.}~\bibnamefont {Kurkjian}}, \bibinfo
  {author} {\bibfnamefont {D.}~\bibnamefont {Kisliuk}}, \bibinfo {author}
  {\bibfnamefont {K.~G.}\ \bibnamefont {Leach}}, \bibinfo {author}
  {\bibfnamefont {E.}~\bibnamefont {McGee}}, \bibinfo {author} {\bibfnamefont
  {J.}~\bibnamefont {Measures}}, \bibinfo {author} {\bibfnamefont
  {D.}~\bibnamefont {M\"ucher}}, \bibinfo {author} {\bibfnamefont
  {J.}~\bibnamefont {Park}}, \bibinfo {author} {\bibfnamefont {F.}~\bibnamefont
  {Sarazin}}, \bibinfo {author} {\bibfnamefont {J.~K.}\ \bibnamefont {Smith}},
  \bibinfo {author} {\bibfnamefont {D.}~\bibnamefont {Southall}}, \bibinfo
  {author} {\bibfnamefont {K.}~\bibnamefont {Starosta}}, \bibinfo {author}
  {\bibfnamefont {C.~E.}\ \bibnamefont {Svensson}}, \bibinfo {author}
  {\bibfnamefont {K.}~\bibnamefont {Whitmore}}, \bibinfo {author}
  {\bibfnamefont {M.}~\bibnamefont {Williams}}, \ and\ \bibinfo {author}
  {\bibfnamefont {C.~Y.}\ \bibnamefont {Wu}},\ }\href {\doibase
  10.1103/PhysRevC.99.051301} {\bibfield  {journal} {\bibinfo  {journal} {Phys.
  Rev. C}\ }\textbf {\bibinfo {volume} {99}},\ \bibinfo {pages} {051301}
  (\bibinfo {year} {2019})}\BibitemShut {NoStop}%
\bibitem [{\citenamefont {Henderson}\ \emph {et~al.}(2018)\citenamefont
  {Henderson} \emph {et~al.}}]{Henderson:2017dqc}%
  \BibitemOpen
  \bibfield  {author} {\bibinfo {author} {\bibfnamefont {J.}~\bibnamefont
  {Henderson}} \emph {et~al.},\ }\href {\doibase
  10.1016/j.physletb.2018.05.064} {\bibfield  {journal} {\bibinfo  {journal}
  {Phys. Lett.}\ }\textbf {\bibinfo {volume} {B782}},\ \bibinfo {pages} {468}
  (\bibinfo {year} {2018})},\ \Eprint {http://arxiv.org/abs/1709.03948}
  {arXiv:1709.03948 [nucl-ex]} \BibitemShut {NoStop}%
\bibitem [{\citenamefont {Williams}\ \emph {et~al.}(2019)\citenamefont
  {Williams}, \citenamefont {Ball}, \citenamefont {Chester}, \citenamefont
  {Domingo}, \citenamefont {Garnsworthy}, \citenamefont {Hackman},
  \citenamefont {Henderson}, \citenamefont {Henderson}, \citenamefont
  {Kr\"ucken}, \citenamefont {Kumar}, \citenamefont {Launey}, \citenamefont
  {Measures}, \citenamefont {Paetkau}, \citenamefont {Park}, \citenamefont
  {Sargsyan}, \citenamefont {Smallcombe}, \citenamefont {Srivastava},
  \citenamefont {Starosta}, \citenamefont {Svensson}, \citenamefont
  {Whitmore},\ and\ \citenamefont {Williams}}]{PhysRevC.100.014322}%
  \BibitemOpen
  \bibfield  {author} {\bibinfo {author} {\bibfnamefont {J.}~\bibnamefont
  {Williams}}, \bibinfo {author} {\bibfnamefont {G.~C.}\ \bibnamefont {Ball}},
  \bibinfo {author} {\bibfnamefont {A.}~\bibnamefont {Chester}}, \bibinfo
  {author} {\bibfnamefont {T.}~\bibnamefont {Domingo}}, \bibinfo {author}
  {\bibfnamefont {A.~B.}\ \bibnamefont {Garnsworthy}}, \bibinfo {author}
  {\bibfnamefont {G.}~\bibnamefont {Hackman}}, \bibinfo {author} {\bibfnamefont
  {J.}~\bibnamefont {Henderson}}, \bibinfo {author} {\bibfnamefont
  {R.}~\bibnamefont {Henderson}}, \bibinfo {author} {\bibfnamefont
  {R.}~\bibnamefont {Kr\"ucken}}, \bibinfo {author} {\bibfnamefont
  {A.}~\bibnamefont {Kumar}}, \bibinfo {author} {\bibfnamefont {K.~D.}\
  \bibnamefont {Launey}}, \bibinfo {author} {\bibfnamefont {J.}~\bibnamefont
  {Measures}}, \bibinfo {author} {\bibfnamefont {O.}~\bibnamefont {Paetkau}},
  \bibinfo {author} {\bibfnamefont {J.}~\bibnamefont {Park}}, \bibinfo {author}
  {\bibfnamefont {G.~H.}\ \bibnamefont {Sargsyan}}, \bibinfo {author}
  {\bibfnamefont {J.}~\bibnamefont {Smallcombe}}, \bibinfo {author}
  {\bibfnamefont {P.~C.}\ \bibnamefont {Srivastava}}, \bibinfo {author}
  {\bibfnamefont {K.}~\bibnamefont {Starosta}}, \bibinfo {author}
  {\bibfnamefont {C.~E.}\ \bibnamefont {Svensson}}, \bibinfo {author}
  {\bibfnamefont {K.}~\bibnamefont {Whitmore}}, \ and\ \bibinfo {author}
  {\bibfnamefont {M.}~\bibnamefont {Williams}},\ }\href {\doibase
  10.1103/PhysRevC.100.014322} {\bibfield  {journal} {\bibinfo  {journal}
  {Phys. Rev. C}\ }\textbf {\bibinfo {volume} {100}},\ \bibinfo {pages}
  {014322} (\bibinfo {year} {2019})}\BibitemShut {NoStop}%
\bibitem [{\citenamefont {Launey}\ \emph {et~al.}(2018)\citenamefont {Launey},
  \citenamefont {Mercenne}, \citenamefont {Sargsyan}, \citenamefont {Shows},
  \citenamefont {Baker}, \citenamefont {Miora}, \citenamefont {Dytrych},\ and\
  \citenamefont {Draayer}}]{LauneySOTANCP42018}%
  \BibitemOpen
  \bibfield  {author} {\bibinfo {author} {\bibfnamefont {K.~D.}\ \bibnamefont
  {Launey}}, \bibinfo {author} {\bibfnamefont {A.}~\bibnamefont {Mercenne}},
  \bibinfo {author} {\bibfnamefont {G.~H.}\ \bibnamefont {Sargsyan}}, \bibinfo
  {author} {\bibfnamefont {H.}~\bibnamefont {Shows}}, \bibinfo {author}
  {\bibfnamefont {R.~B.}\ \bibnamefont {Baker}}, \bibinfo {author}
  {\bibfnamefont {M.~E.}\ \bibnamefont {Miora}}, \bibinfo {author}
  {\bibfnamefont {T.}~\bibnamefont {Dytrych}}, \ and\ \bibinfo {author}
  {\bibfnamefont {J.~P.}\ \bibnamefont {Draayer}},\ }in\ \href@noop {} {\emph
  {\bibinfo {booktitle} {{Proceedings of the 4th International Workshop on
  'State of the Art in Nuclear Cluster Physics' (SOTANCP4), May 2018,
  Galveston, Texas}}}},\ Vol.\ \bibinfo {volume} {2038}\ (\bibinfo  {publisher}
  {{AIP Conference Proceedings}},\ \bibinfo {year} {2018})\BibitemShut
  {NoStop}%
\bibitem [{\citenamefont {Launey}\ \emph {et~al.}(2021)\citenamefont {Launey},
  \citenamefont {Mercenne},\ and\ \citenamefont {Dytrych}}]{LauneyMD_ARNPS21}%
  \BibitemOpen
  \bibfield  {author} {\bibinfo {author} {\bibfnamefont {K.~D.}\ \bibnamefont
  {Launey}}, \bibinfo {author} {\bibfnamefont {A.}~\bibnamefont {Mercenne}}, \
  and\ \bibinfo {author} {\bibfnamefont {T.}~\bibnamefont {Dytrych}},\ }\href
  {\doibase 10.1146/annurev-nucl-102419-033316} {\bibfield  {journal} {\bibinfo
   {journal} {Annu. Rev. Nucl. Part. Sci.}\ }\textbf {\bibinfo {volume} {71}},\
  \bibinfo {pages} {253} (\bibinfo {year} {2021})}\BibitemShut {NoStop}%
\bibitem [{\citenamefont {Johnson}\ \emph {et~al.}(2020)\citenamefont
  {Johnson}, \citenamefont {Launey} \emph {et~al.}}]{FRIBTAwhite2018}%
  \BibitemOpen
  \bibfield  {author} {\bibinfo {author} {\bibfnamefont {C.~W.}\ \bibnamefont
  {Johnson}}, \bibinfo {author} {\bibfnamefont {K.~D.}\ \bibnamefont {Launey}},
   \emph {et~al.},\ }\href@noop {} {\bibfield  {journal} {\bibinfo  {journal}
  {J. Phys. G}\ }\textbf {\bibinfo {volume} {47}},\ \bibinfo {pages} {23001}
  (\bibinfo {year} {2020})},\ \bibinfo {note} {arXiv:1912.00451}\BibitemShut
  {NoStop}%
\bibitem [{\citenamefont {Quaglioni}\ and\ \citenamefont
  {Navrátil}(2020)}]{doi:10.1080/10619127.2020.1752089}%
  \BibitemOpen
  \bibfield  {author} {\bibinfo {author} {\bibfnamefont {S.}~\bibnamefont
  {Quaglioni}}\ and\ \bibinfo {author} {\bibfnamefont {P.}~\bibnamefont
  {Navrátil}},\ }\href {\doibase 10.1080/10619127.2020.1752089} {\bibfield
  {journal} {\bibinfo  {journal} {Nuclear Physics News}\ }\textbf {\bibinfo
  {volume} {30}},\ \bibinfo {pages} {12} (\bibinfo {year} {2020})}\BibitemShut
  {NoStop}%
\bibitem [{\citenamefont {Nollett}\ \emph {et~al.}(2007)\citenamefont
  {Nollett}, \citenamefont {Pieper}, \citenamefont {Wiringa}, \citenamefont
  {Carlson},\ and\ \citenamefont {Hale}}]{NollettPWCH07}%
  \BibitemOpen
  \bibfield  {author} {\bibinfo {author} {\bibfnamefont {K.}~\bibnamefont
  {Nollett}}, \bibinfo {author} {\bibfnamefont {S.}~\bibnamefont {Pieper}},
  \bibinfo {author} {\bibfnamefont {R.}~\bibnamefont {Wiringa}}, \bibinfo
  {author} {\bibfnamefont {J.}~\bibnamefont {Carlson}}, \ and\ \bibinfo
  {author} {\bibfnamefont {G.}~\bibnamefont {Hale}},\ }\href@noop {} {\bibfield
   {journal} {\bibinfo  {journal} {Phys. Rev. Lett.}\ }\textbf {\bibinfo
  {volume} {99}},\ \bibinfo {pages} {022502} (\bibinfo {year}
  {2007})}\BibitemShut {NoStop}%
\bibitem [{\citenamefont {Hagen}\ \emph {et~al.}(2007)\citenamefont {Hagen},
  \citenamefont {Dean}, \citenamefont {Hjorth-Jensen},\ and\ \citenamefont
  {Papenbrock}}]{HagenDHP07}%
  \BibitemOpen
  \bibfield  {author} {\bibinfo {author} {\bibfnamefont {G.}~\bibnamefont
  {Hagen}}, \bibinfo {author} {\bibfnamefont {D.}~\bibnamefont {Dean}},
  \bibinfo {author} {\bibfnamefont {M.}~\bibnamefont {Hjorth-Jensen}}, \ and\
  \bibinfo {author} {\bibfnamefont {T.}~\bibnamefont {Papenbrock}},\
  }\href@noop {} {\bibfield  {journal} {\bibinfo  {journal} {Phys. Lett. B}\
  }\textbf {\bibinfo {volume} {656}},\ \bibinfo {pages} {169} (\bibinfo {year}
  {2007})}\BibitemShut {NoStop}%
\bibitem [{\citenamefont {Quaglioni}\ and\ \citenamefont
  {Navr\'atil}(2008{\natexlab{a}})}]{PhysRevLett.101.092501}%
  \BibitemOpen
  \bibfield  {author} {\bibinfo {author} {\bibfnamefont {S.}~\bibnamefont
  {Quaglioni}}\ and\ \bibinfo {author} {\bibfnamefont {P.}~\bibnamefont
  {Navr\'atil}},\ }\href {\doibase 10.1103/PhysRevLett.101.092501} {\bibfield
  {journal} {\bibinfo  {journal} {Phys. Rev. Lett.}\ }\textbf {\bibinfo
  {volume} {101}},\ \bibinfo {pages} {092501} (\bibinfo {year}
  {2008}{\natexlab{a}})}\BibitemShut {NoStop}%
\bibitem [{\citenamefont {Elhatisari}\ \emph {et~al.}(2015)\citenamefont
  {Elhatisari}, \citenamefont {Lee}, \citenamefont {Rupak}, \citenamefont
  {Epelbaum} \emph {et~al.}}]{ElhatisariLRE15}%
  \BibitemOpen
  \bibfield  {author} {\bibinfo {author} {\bibfnamefont {S.}~\bibnamefont
  {Elhatisari}}, \bibinfo {author} {\bibfnamefont {D.}~\bibnamefont {Lee}},
  \bibinfo {author} {\bibfnamefont {G.}~\bibnamefont {Rupak}}, \bibinfo
  {author} {\bibfnamefont {E.}~\bibnamefont {Epelbaum}},  \emph {et~al.},\
  }\href@noop {} {\bibfield  {journal} {\bibinfo  {journal} {Nature}\ }\textbf
  {\bibinfo {volume} {528}},\ \bibinfo {pages} {111} (\bibinfo {year}
  {2015})}\BibitemShut {NoStop}%
\bibitem [{\citenamefont {Zhang}\ \emph {et~al.}(2020)\citenamefont {Zhang},
  \citenamefont {Stroberg}, \citenamefont {Navr\'atil}, \citenamefont {Gwak},
  \citenamefont {Melendez}, \citenamefont {Furnstahl},\ and\ \citenamefont
  {Holt}}]{PhysRevLett.125.112503}%
  \BibitemOpen
  \bibfield  {author} {\bibinfo {author} {\bibfnamefont {X.}~\bibnamefont
  {Zhang}}, \bibinfo {author} {\bibfnamefont {S.~R.}\ \bibnamefont {Stroberg}},
  \bibinfo {author} {\bibfnamefont {P.}~\bibnamefont {Navr\'atil}}, \bibinfo
  {author} {\bibfnamefont {C.}~\bibnamefont {Gwak}}, \bibinfo {author}
  {\bibfnamefont {J.~A.}\ \bibnamefont {Melendez}}, \bibinfo {author}
  {\bibfnamefont {R.~J.}\ \bibnamefont {Furnstahl}}, \ and\ \bibinfo {author}
  {\bibfnamefont {J.~D.}\ \bibnamefont {Holt}},\ }\href {\doibase
  10.1103/PhysRevLett.125.112503} {\bibfield  {journal} {\bibinfo  {journal}
  {Phys. Rev. Lett.}\ }\textbf {\bibinfo {volume} {125}},\ \bibinfo {pages}
  {112503} (\bibinfo {year} {2020})}\BibitemShut {NoStop}%
\bibitem [{\citenamefont {Mercenne}\ \emph {et~al.}(2019)\citenamefont
  {Mercenne}, \citenamefont {Michel},\ and\ \citenamefont
  {P{\l}oszajczak}}]{mercenne_2019}%
  \BibitemOpen
  \bibfield  {author} {\bibinfo {author} {\bibfnamefont {A.}~\bibnamefont
  {Mercenne}}, \bibinfo {author} {\bibfnamefont {N.}~\bibnamefont {Michel}}, \
  and\ \bibinfo {author} {\bibfnamefont {M.}~\bibnamefont {P{\l}oszajczak}},\
  }\href@noop {} {\bibfield  {journal} {\bibinfo  {journal} {Phys. Rev. C}\
  }\textbf {\bibinfo {volume} {99}},\ \bibinfo {pages} {044606} (\bibinfo
  {year} {2019})}\BibitemShut {NoStop}%
\bibitem [{\citenamefont {Burrows}\ \emph {et~al.}(2020)\citenamefont
  {Burrows}, \citenamefont {Baker}, \citenamefont {Elster}, \citenamefont
  {Weppner}, \citenamefont {Launey}, \citenamefont {Maris},\ and\ \citenamefont
  {Popa}}]{BurrowsBEWLMP20}%
  \BibitemOpen
  \bibfield  {author} {\bibinfo {author} {\bibfnamefont {M.}~\bibnamefont
  {Burrows}}, \bibinfo {author} {\bibfnamefont {R.~B.}\ \bibnamefont {Baker}},
  \bibinfo {author} {\bibfnamefont {C.}~\bibnamefont {Elster}}, \bibinfo
  {author} {\bibfnamefont {S.~P.}\ \bibnamefont {Weppner}}, \bibinfo {author}
  {\bibfnamefont {K.~D.}\ \bibnamefont {Launey}}, \bibinfo {author}
  {\bibfnamefont {P.}~\bibnamefont {Maris}}, \ and\ \bibinfo {author}
  {\bibfnamefont {G.}~\bibnamefont {Popa}},\ }\href {\doibase
  10.1103/PhysRevC.102.034606} {\bibfield  {journal} {\bibinfo  {journal}
  {Phys. Rev. C}\ }\textbf {\bibinfo {volume} {102}},\ \bibinfo {pages}
  {034606} (\bibinfo {year} {2020})}\BibitemShut {NoStop}%
\bibitem [{\citenamefont {Bacca}\ \emph {et~al.}(2014)\citenamefont {Bacca},
  \citenamefont {Barnea}, \citenamefont {Hagen}, \citenamefont {Miorelli},
  \citenamefont {Orlandini},\ and\ \citenamefont
  {Papenbrock}}]{PhysRevC.90.064619}%
  \BibitemOpen
  \bibfield  {author} {\bibinfo {author} {\bibfnamefont {S.}~\bibnamefont
  {Bacca}}, \bibinfo {author} {\bibfnamefont {N.}~\bibnamefont {Barnea}},
  \bibinfo {author} {\bibfnamefont {G.}~\bibnamefont {Hagen}}, \bibinfo
  {author} {\bibfnamefont {M.}~\bibnamefont {Miorelli}}, \bibinfo {author}
  {\bibfnamefont {G.}~\bibnamefont {Orlandini}}, \ and\ \bibinfo {author}
  {\bibfnamefont {T.}~\bibnamefont {Papenbrock}},\ }\href {\doibase
  10.1103/PhysRevC.90.064619} {\bibfield  {journal} {\bibinfo  {journal} {Phys.
  Rev. C}\ }\textbf {\bibinfo {volume} {90}},\ \bibinfo {pages} {064619}
  (\bibinfo {year} {2014})}\BibitemShut {NoStop}%
\bibitem [{\citenamefont {Navr\'{a}til}\ and\ \citenamefont
  {Quaglioni}(2012)}]{NavratilQ12}%
  \BibitemOpen
  \bibfield  {author} {\bibinfo {author} {\bibfnamefont {P.}~\bibnamefont
  {Navr\'{a}til}}\ and\ \bibinfo {author} {\bibfnamefont {S.}~\bibnamefont
  {Quaglioni}},\ }\href@noop {} {\bibfield  {journal} {\bibinfo  {journal}
  {Phys. Rev. Lett.}\ }\textbf {\bibinfo {volume} {108}},\ \bibinfo {pages}
  {042503} (\bibinfo {year} {2012})}\BibitemShut {NoStop}%
\bibitem [{\citenamefont {Girlanda}\ \emph {et~al.}(2010)\citenamefont
  {Girlanda}, \citenamefont {Kievsky}, \citenamefont {Marcucci}, \citenamefont
  {Pastore}, \citenamefont {Schiavilla},\ and\ \citenamefont
  {Viviani}}]{PhysRevLett.105.232502}%
  \BibitemOpen
  \bibfield  {author} {\bibinfo {author} {\bibfnamefont {L.}~\bibnamefont
  {Girlanda}}, \bibinfo {author} {\bibfnamefont {A.}~\bibnamefont {Kievsky}},
  \bibinfo {author} {\bibfnamefont {L.~E.}\ \bibnamefont {Marcucci}}, \bibinfo
  {author} {\bibfnamefont {S.}~\bibnamefont {Pastore}}, \bibinfo {author}
  {\bibfnamefont {R.}~\bibnamefont {Schiavilla}}, \ and\ \bibinfo {author}
  {\bibfnamefont {M.}~\bibnamefont {Viviani}},\ }\href {\doibase
  10.1103/PhysRevLett.105.232502} {\bibfield  {journal} {\bibinfo  {journal}
  {Phys. Rev. Lett.}\ }\textbf {\bibinfo {volume} {105}},\ \bibinfo {pages}
  {232502} (\bibinfo {year} {2010})}\BibitemShut {NoStop}%
\bibitem [{\citenamefont {Kravvaris}\ and\ \citenamefont
  {Volya}(2017)}]{kravvaris_2017}%
  \BibitemOpen
  \bibfield  {author} {\bibinfo {author} {\bibfnamefont {K.}~\bibnamefont
  {Kravvaris}}\ and\ \bibinfo {author} {\bibfnamefont {A.}~\bibnamefont
  {Volya}},\ }\href@noop {} {\bibfield  {journal} {\bibinfo  {journal} {Phys.
  Rev. Lett.}\ }\textbf {\bibinfo {volume} {119}},\ \bibinfo {pages} {062501}
  (\bibinfo {year} {2017})}\BibitemShut {NoStop}%
\bibitem [{\citenamefont {Dreyfuss}\ \emph {et~al.}(2020)\citenamefont
  {Dreyfuss}, \citenamefont {Launey}, \citenamefont {Escher}, \citenamefont
  {Sargsyan}, \citenamefont {Baker}, \citenamefont {Dytrych},\ and\
  \citenamefont {Draayer}}]{DreyfussLESBDD20}%
  \BibitemOpen
  \bibfield  {author} {\bibinfo {author} {\bibfnamefont {A.~C.}\ \bibnamefont
  {Dreyfuss}}, \bibinfo {author} {\bibfnamefont {K.~D.}\ \bibnamefont
  {Launey}}, \bibinfo {author} {\bibfnamefont {J.~E.}\ \bibnamefont {Escher}},
  \bibinfo {author} {\bibfnamefont {G.~H.}\ \bibnamefont {Sargsyan}}, \bibinfo
  {author} {\bibfnamefont {R.~B.}\ \bibnamefont {Baker}}, \bibinfo {author}
  {\bibfnamefont {T.}~\bibnamefont {Dytrych}}, \ and\ \bibinfo {author}
  {\bibfnamefont {J.~P.}\ \bibnamefont {Draayer}},\ }\href {\doibase
  10.1103/PhysRevC.102.044608} {\bibfield  {journal} {\bibinfo  {journal}
  {Phys. Rev. C}\ }\textbf {\bibinfo {volume} {102}},\ \bibinfo {pages}
  {044608} (\bibinfo {year} {2020})}\BibitemShut {NoStop}%
\bibitem [{\citenamefont {Lazauskas}\ \emph {et~al.}(2019)\citenamefont
  {Lazauskas}, \citenamefont {Hiyama},\ and\ \citenamefont
  {Carbonnell}}]{lazauskas_2019}%
  \BibitemOpen
  \bibfield  {author} {\bibinfo {author} {\bibfnamefont {R.}~\bibnamefont
  {Lazauskas}}, \bibinfo {author} {\bibfnamefont {E.}~\bibnamefont {Hiyama}}, \
  and\ \bibinfo {author} {\bibfnamefont {J.}~\bibnamefont {Carbonnell}},\
  }\href@noop {} {\bibfield  {journal} {\bibinfo  {journal} {Phys. Lett. B}\
  }\textbf {\bibinfo {volume} {791}},\ \bibinfo {pages} {335} (\bibinfo {year}
  {2019})}\BibitemShut {NoStop}%
\bibitem [{\citenamefont {Hupin}\ \emph {et~al.}(2019)\citenamefont {Hupin},
  \citenamefont {Quaglioni},\ and\ \citenamefont {Navr\'{a}til}}]{HupinQN19}%
  \BibitemOpen
  \bibfield  {author} {\bibinfo {author} {\bibfnamefont {G.}~\bibnamefont
  {Hupin}}, \bibinfo {author} {\bibfnamefont {S.}~\bibnamefont {Quaglioni}}, \
  and\ \bibinfo {author} {\bibfnamefont {P.}~\bibnamefont {Navr\'{a}til}},\
  }\href@noop {} {\bibfield  {journal} {\bibinfo  {journal} {Nature
  Communications}\ }\textbf {\bibinfo {volume} {10}},\ \bibinfo {pages} {351}
  (\bibinfo {year} {2019})}\BibitemShut {NoStop}%
\bibitem [{\citenamefont {Bedaque}\ and\ \citenamefont {van
  Kolck}(2002)}]{BedaqueVKolck02}%
  \BibitemOpen
  \bibfield  {author} {\bibinfo {author} {\bibfnamefont {P.~F.}\ \bibnamefont
  {Bedaque}}\ and\ \bibinfo {author} {\bibfnamefont {U.}~\bibnamefont {van
  Kolck}},\ }\href {\doibase 10.1146/annurev.nucl.52.050102.090637} {\bibfield
  {journal} {\bibinfo  {journal} {Annu. Rev. Nucl. Part. Sci.}\ }\textbf
  {\bibinfo {volume} {52}},\ \bibinfo {pages} {339} (\bibinfo {year}
  {2002})}\BibitemShut {NoStop}%
\bibitem [{\citenamefont {Epelbaum}\ \emph {et~al.}(2002)\citenamefont
  {Epelbaum}, \citenamefont {Nogga}, \citenamefont {Gl{\"o}ckle}, \citenamefont
  {Kamada}, \citenamefont {Mei{\ss}ner},\ and\ \citenamefont
  {Witala}}]{EpelbaumNGKMW02}%
  \BibitemOpen
  \bibfield  {author} {\bibinfo {author} {\bibfnamefont {E.}~\bibnamefont
  {Epelbaum}}, \bibinfo {author} {\bibfnamefont {A.}~\bibnamefont {Nogga}},
  \bibinfo {author} {\bibfnamefont {W.}~\bibnamefont {Gl{\"o}ckle}}, \bibinfo
  {author} {\bibfnamefont {H.}~\bibnamefont {Kamada}}, \bibinfo {author}
  {\bibfnamefont {U.-G.}\ \bibnamefont {Mei{\ss}ner}}, \ and\ \bibinfo {author}
  {\bibfnamefont {H.}~\bibnamefont {Witala}},\ }\href@noop {} {\bibfield
  {journal} {\bibinfo  {journal} {Phys. Rev. C}\ }\textbf {\bibinfo {volume}
  {66}},\ \bibinfo {pages} {064001} (\bibinfo {year} {2002})}\BibitemShut
  {NoStop}%
\bibitem [{\citenamefont {Entem}\ and\ \citenamefont
  {Machleidt}(2003)}]{EntemM03}%
  \BibitemOpen
  \bibfield  {author} {\bibinfo {author} {\bibfnamefont {D.~R.}\ \bibnamefont
  {Entem}}\ and\ \bibinfo {author} {\bibfnamefont {R.}~\bibnamefont
  {Machleidt}},\ }\href@noop {} {\bibfield  {journal} {\bibinfo  {journal}
  {Phys. Rev. C}\ }\textbf {\bibinfo {volume} {68}},\ \bibinfo {pages} {041001}
  (\bibinfo {year} {2003})}\BibitemShut {NoStop}%
\bibitem [{\citenamefont {Epelbaum}(2006)}]{Epelbaum06}%
  \BibitemOpen
  \bibfield  {author} {\bibinfo {author} {\bibfnamefont {E.}~\bibnamefont
  {Epelbaum}},\ }\href@noop {} {\bibfield  {journal} {\bibinfo  {journal}
  {Prog. Part. Nucl. Phys.}\ }\textbf {\bibinfo {volume} {57}},\ \bibinfo
  {pages} {654} (\bibinfo {year} {2006})}\BibitemShut {NoStop}%
\bibitem [{\citenamefont {Wildermuth}\ and\ \citenamefont
  {Tang}(1977)}]{WildermuthT77}%
  \BibitemOpen
  \bibfield  {author} {\bibinfo {author} {\bibfnamefont {K.}~\bibnamefont
  {Wildermuth}}\ and\ \bibinfo {author} {\bibfnamefont {Y.}~\bibnamefont
  {Tang}},\ }\href@noop {} {\emph {\bibinfo {title} {{A Unified Theory of The
  Nucleus}}}}\ (\bibinfo  {publisher} {Braunschweig, Vieweg},\ \bibinfo {year}
  {1977})\BibitemShut {NoStop}%
\bibitem [{\citenamefont {Tang}\ \emph {et~al.}(1978)\citenamefont {Tang},
  \citenamefont {LeMere},\ and\ \citenamefont {Thompson}}]{RGM_tang_1978}%
  \BibitemOpen
  \bibfield  {author} {\bibinfo {author} {\bibfnamefont {Y.~C.}\ \bibnamefont
  {Tang}}, \bibinfo {author} {\bibfnamefont {M.}~\bibnamefont {LeMere}}, \ and\
  \bibinfo {author} {\bibfnamefont {D.~R.}\ \bibnamefont {Thompson}},\
  }\href@noop {} {\bibfield  {journal} {\bibinfo  {journal} {Phys. Rep.}\
  }\textbf {\bibinfo {volume} {47}},\ \bibinfo {pages} {167} (\bibinfo {year}
  {1978})}\BibitemShut {NoStop}%
\bibitem [{\citenamefont {Elliott}(1958{\natexlab{a}})}]{Elliott58}%
  \BibitemOpen
  \bibfield  {author} {\bibinfo {author} {\bibfnamefont {J.~P.}\ \bibnamefont
  {Elliott}},\ }\href@noop {} {\bibfield  {journal} {\bibinfo  {journal} {Proc.
  Roy. Soc. A}\ }\textbf {\bibinfo {volume} {245}},\ \bibinfo {pages} {128}
  (\bibinfo {year} {1958}{\natexlab{a}})}\BibitemShut {NoStop}%
\bibitem [{\citenamefont {Elliott}(1958{\natexlab{b}})}]{Elliott58b}%
  \BibitemOpen
  \bibfield  {author} {\bibinfo {author} {\bibfnamefont {J.~P.}\ \bibnamefont
  {Elliott}},\ }\href@noop {} {\bibfield  {journal} {\bibinfo  {journal} {Proc.
  Roy. Soc. A}\ }\textbf {\bibinfo {volume} {245}},\ \bibinfo {pages} {562}
  (\bibinfo {year} {1958}{\natexlab{b}})}\BibitemShut {NoStop}%
\bibitem [{\citenamefont {Hecht}(1977)}]{Hecht77_NPA283}%
  \BibitemOpen
  \bibfield  {author} {\bibinfo {author} {\bibfnamefont {K.~T.}\ \bibnamefont
  {Hecht}},\ }\href@noop {} {\bibfield  {journal} {\bibinfo  {journal} {Nucl.
  Phys. A}\ }\textbf {\bibinfo {volume} {283}},\ \bibinfo {pages} {223}
  (\bibinfo {year} {1977})}\BibitemShut {NoStop}%
\bibitem [{\citenamefont {Hecht}\ and\ \citenamefont
  {Zahn}(1978)}]{Hecht_1978}%
  \BibitemOpen
  \bibfield  {author} {\bibinfo {author} {\bibfnamefont {K.~T.}\ \bibnamefont
  {Hecht}}\ and\ \bibinfo {author} {\bibfnamefont {W.}~\bibnamefont {Zahn}},\
  }\href@noop {} {\bibfield  {journal} {\bibinfo  {journal} {Nucl. Phys. A}\
  }\textbf {\bibinfo {volume} {318}},\ \bibinfo {pages} {1} (\bibinfo {year}
  {1978})}\BibitemShut {NoStop}%
\bibitem [{\citenamefont {Hecht}\ and\ \citenamefont
  {Suzuki}(1982)}]{HechtS82}%
  \BibitemOpen
  \bibfield  {author} {\bibinfo {author} {\bibfnamefont {K.~T.}\ \bibnamefont
  {Hecht}}\ and\ \bibinfo {author} {\bibfnamefont {Y.}~\bibnamefont {Suzuki}},\
  }\href@noop {} {\bibfield  {journal} {\bibinfo  {journal} {J. Math. Phys.}\
  }\textbf {\bibinfo {volume} {24}},\ \bibinfo {pages} {785} (\bibinfo {year}
  {1982})}\BibitemShut {NoStop}%
\bibitem [{\citenamefont {Hecht}\ \emph {et~al.}(1981)\citenamefont {Hecht},
  \citenamefont {Reske}, \citenamefont {Seligman},\ and\ \citenamefont
  {Zahn}}]{HechtRSZ81}%
  \BibitemOpen
  \bibfield  {author} {\bibinfo {author} {\bibfnamefont {K.~T.}\ \bibnamefont
  {Hecht}}, \bibinfo {author} {\bibfnamefont {E.~J.}\ \bibnamefont {Reske}},
  \bibinfo {author} {\bibfnamefont {T.~H.}\ \bibnamefont {Seligman}}, \ and\
  \bibinfo {author} {\bibfnamefont {W.}~\bibnamefont {Zahn}},\ }\href@noop {}
  {\bibfield  {journal} {\bibinfo  {journal} {Nucl. Phys. A}\ }\textbf
  {\bibinfo {volume} {356}},\ \bibinfo {pages} {146} (\bibinfo {year}
  {1981})}\BibitemShut {NoStop}%
\bibitem [{\citenamefont {Hecht}\ and\ \citenamefont
  {Braunschweig}(1978)}]{HechtB82}%
  \BibitemOpen
  \bibfield  {author} {\bibinfo {author} {\bibfnamefont {K.~T.}\ \bibnamefont
  {Hecht}}\ and\ \bibinfo {author} {\bibfnamefont {D.}~\bibnamefont
  {Braunschweig}},\ }\href@noop {} {\bibfield  {journal} {\bibinfo  {journal}
  {Nucl. Phys. A}\ }\textbf {\bibinfo {volume} {295}},\ \bibinfo {pages} {34}
  (\bibinfo {year} {1978})}\BibitemShut {NoStop}%
\bibitem [{\citenamefont {Suzuki}\ and\ \citenamefont
  {Hecht}(1982)}]{SuzukiH82}%
  \BibitemOpen
  \bibfield  {author} {\bibinfo {author} {\bibfnamefont {Y.}~\bibnamefont
  {Suzuki}}\ and\ \bibinfo {author} {\bibfnamefont {K.~T.}\ \bibnamefont
  {Hecht}},\ }\href@noop {} {\bibfield  {journal} {\bibinfo  {journal} {Nucl.
  Phys. A}\ }\textbf {\bibinfo {volume} {388}},\ \bibinfo {pages} {102}
  (\bibinfo {year} {1982})}\BibitemShut {NoStop}%
\bibitem [{\citenamefont {Suzuki}(1986)}]{Suzuki86}%
  \BibitemOpen
  \bibfield  {author} {\bibinfo {author} {\bibfnamefont {Y.}~\bibnamefont
  {Suzuki}},\ }\href@noop {} {\bibfield  {journal} {\bibinfo  {journal} {Nucl.
  Phys. A}\ }\textbf {\bibinfo {volume} {448}},\ \bibinfo {pages} {395}
  (\bibinfo {year} {1986})}\BibitemShut {NoStop}%
\bibitem [{\citenamefont {Suzuki}\ and\ \citenamefont
  {Hecht}(1986)}]{SuzukiH86}%
  \BibitemOpen
  \bibfield  {author} {\bibinfo {author} {\bibfnamefont {Y.}~\bibnamefont
  {Suzuki}}\ and\ \bibinfo {author} {\bibfnamefont {K.~T.}\ \bibnamefont
  {Hecht}},\ }\href@noop {} {\bibfield  {journal} {\bibinfo  {journal} {Nucl.
  Phys. A}\ }\textbf {\bibinfo {volume} {455}},\ \bibinfo {pages} {315}
  (\bibinfo {year} {1986})}\BibitemShut {NoStop}%
\bibitem [{\citenamefont {Suzuki}\ and\ \citenamefont
  {Hecht}(1987)}]{SuzukiH87}%
  \BibitemOpen
  \bibfield  {author} {\bibinfo {author} {\bibfnamefont {Y.}~\bibnamefont
  {Suzuki}}\ and\ \bibinfo {author} {\bibfnamefont {K.~T.}\ \bibnamefont
  {Hecht}},\ }\href@noop {} {\bibfield  {journal} {\bibinfo  {journal} {Prog.
  Theor. Phys.}\ }\textbf {\bibinfo {volume} {77}},\ \bibinfo {pages} {190}
  (\bibinfo {year} {1987})}\BibitemShut {NoStop}%
\bibitem [{\citenamefont {Suzuki}(1987)}]{Suzuki87}%
  \BibitemOpen
  \bibfield  {author} {\bibinfo {author} {\bibfnamefont {Y.}~\bibnamefont
  {Suzuki}},\ }\href@noop {} {\bibfield  {journal} {\bibinfo  {journal} {Nucl.
  Phys. A}\ }\textbf {\bibinfo {volume} {470}},\ \bibinfo {pages} {119}
  (\bibinfo {year} {1987})}\BibitemShut {NoStop}%
\bibitem [{\citenamefont {Suzuki}\ and\ \citenamefont
  {Hara}(1989)}]{SuzukiH89}%
  \BibitemOpen
  \bibfield  {author} {\bibinfo {author} {\bibfnamefont {Y.}~\bibnamefont
  {Suzuki}}\ and\ \bibinfo {author} {\bibfnamefont {S.}~\bibnamefont {Hara}},\
  }\href@noop {} {\bibfield  {journal} {\bibinfo  {journal} {Phys. Rev. C}\
  }\textbf {\bibinfo {volume} {39}},\ \bibinfo {pages} {658} (\bibinfo {year}
  {1989})}\BibitemShut {NoStop}%
\bibitem [{\citenamefont {Suzuki}(1976{\natexlab{a}})}]{Suzuki76a}%
  \BibitemOpen
  \bibfield  {author} {\bibinfo {author} {\bibfnamefont {Y.}~\bibnamefont
  {Suzuki}},\ }\href@noop {} {\bibfield  {journal} {\bibinfo  {journal} {Prog.
  Theor. Phys.}\ }\textbf {\bibinfo {volume} {55}},\ \bibinfo {pages} {1751}
  (\bibinfo {year} {1976}{\natexlab{a}})}\BibitemShut {NoStop}%
\bibitem [{\citenamefont {Suzuki}(1976{\natexlab{b}})}]{Suzuki76b}%
  \BibitemOpen
  \bibfield  {author} {\bibinfo {author} {\bibfnamefont {Y.}~\bibnamefont
  {Suzuki}},\ }\href@noop {} {\bibfield  {journal} {\bibinfo  {journal} {Prog.
  Theor. Phys.}\ }\textbf {\bibinfo {volume} {56}},\ \bibinfo {pages} {111}
  (\bibinfo {year} {1976}{\natexlab{b}})}\BibitemShut {NoStop}%
\bibitem [{\citenamefont {Suzuki}\ \emph {et~al.}(2003)\citenamefont {Suzuki},
  \citenamefont {Lovas}, \citenamefont {Yabana},\ and\ \citenamefont
  {Varga}}]{SuzukiLYV03}%
  \BibitemOpen
  \bibfield  {author} {\bibinfo {author} {\bibfnamefont {Y.}~\bibnamefont
  {Suzuki}}, \bibinfo {author} {\bibfnamefont {R.~G.}\ \bibnamefont {Lovas}},
  \bibinfo {author} {\bibfnamefont {K.}~\bibnamefont {Yabana}}, \ and\ \bibinfo
  {author} {\bibfnamefont {K.}~\bibnamefont {Varga}},\ }\href@noop {} {\emph
  {\bibinfo {title} {Structure and reactions of exotic nuclei}}}\ (\bibinfo
  {publisher} {Taylor \& Francis, London and New York},\ \bibinfo {year}
  {2003})\BibitemShut {NoStop}%
\bibitem [{\citenamefont {Navr\'{a}til}\ \emph {et~al.}(2009)\citenamefont
  {Navr\'{a}til}, \citenamefont {Quaglioni}, \citenamefont {Stetcu},\ and\
  \citenamefont {Barrett}}]{NavratilQSB09}%
  \BibitemOpen
  \bibfield  {author} {\bibinfo {author} {\bibfnamefont {P.}~\bibnamefont
  {Navr\'{a}til}}, \bibinfo {author} {\bibfnamefont {S.}~\bibnamefont
  {Quaglioni}}, \bibinfo {author} {\bibfnamefont {I.}~\bibnamefont {Stetcu}}, \
  and\ \bibinfo {author} {\bibfnamefont {B.~R.}\ \bibnamefont {Barrett}},\
  }\href@noop {} {\bibfield  {journal} {\bibinfo  {journal} {J. Phys. G: Nucl.
  Part.}\ }\textbf {\bibinfo {volume} {36}},\ \bibinfo {pages} {083101}
  (\bibinfo {year} {2009})}\BibitemShut {NoStop}%
\bibitem [{\citenamefont {Barrett}\ \emph {et~al.}(2013)\citenamefont
  {Barrett}, \citenamefont {Navr\'{a}til},\ and\ \citenamefont
  {Vary}}]{BarrettNV13}%
  \BibitemOpen
  \bibfield  {author} {\bibinfo {author} {\bibfnamefont {B.}~\bibnamefont
  {Barrett}}, \bibinfo {author} {\bibfnamefont {P.}~\bibnamefont
  {Navr\'{a}til}}, \ and\ \bibinfo {author} {\bibfnamefont {J.}~\bibnamefont
  {Vary}},\ }\href@noop {} {\bibfield  {journal} {\bibinfo  {journal} {Prog.
  Part. Nucl. Phys.}\ }\textbf {\bibinfo {volume} {69}},\ \bibinfo {pages}
  {131} (\bibinfo {year} {2013})}\BibitemShut {NoStop}%
\bibitem [{\citenamefont {Quaglioni}\ and\ \citenamefont
  {Navr\'atil}(2008{\natexlab{b}})}]{quaglioni_2008}%
  \BibitemOpen
  \bibfield  {author} {\bibinfo {author} {\bibfnamefont {S.}~\bibnamefont
  {Quaglioni}}\ and\ \bibinfo {author} {\bibfnamefont {P.}~\bibnamefont
  {Navr\'atil}},\ }\href@noop {} {\bibfield  {journal} {\bibinfo  {journal}
  {Phys. Rev. Lett.}\ }\textbf {\bibinfo {volume} {101}},\ \bibinfo {pages}
  {092501} (\bibinfo {year} {2008}{\natexlab{b}})}\BibitemShut {NoStop}%
\bibitem [{\citenamefont {Quaglioni}\ and\ \citenamefont
  {Navr\'{a}til}(2009)}]{QuaglioniN09}%
  \BibitemOpen
  \bibfield  {author} {\bibinfo {author} {\bibfnamefont {S.}~\bibnamefont
  {Quaglioni}}\ and\ \bibinfo {author} {\bibfnamefont {P.}~\bibnamefont
  {Navr\'{a}til}},\ }\href@noop {} {\bibfield  {journal} {\bibinfo  {journal}
  {Phys. Rev. C}\ }\textbf {\bibinfo {volume} {79}},\ \bibinfo {pages} {044606}
  (\bibinfo {year} {2009})}\BibitemShut {NoStop}%
\bibitem [{\citenamefont {Hupin}\ \emph {et~al.}(2013)\citenamefont {Hupin},
  \citenamefont {Langhammer}, \citenamefont {Navr\'atil}, \citenamefont
  {Quaglioni}, \citenamefont {Calci},\ and\ \citenamefont
  {Roth}}]{PhysRevC.88.054622}%
  \BibitemOpen
  \bibfield  {author} {\bibinfo {author} {\bibfnamefont {G.}~\bibnamefont
  {Hupin}}, \bibinfo {author} {\bibfnamefont {J.}~\bibnamefont {Langhammer}},
  \bibinfo {author} {\bibfnamefont {P.}~\bibnamefont {Navr\'atil}}, \bibinfo
  {author} {\bibfnamefont {S.}~\bibnamefont {Quaglioni}}, \bibinfo {author}
  {\bibfnamefont {A.}~\bibnamefont {Calci}}, \ and\ \bibinfo {author}
  {\bibfnamefont {R.}~\bibnamefont {Roth}},\ }\href {\doibase
  10.1103/PhysRevC.88.054622} {\bibfield  {journal} {\bibinfo  {journal} {Phys.
  Rev. C}\ }\textbf {\bibinfo {volume} {88}},\ \bibinfo {pages} {054622}
  (\bibinfo {year} {2013})}\BibitemShut {NoStop}%
\bibitem [{\citenamefont {Baroni}\ \emph
  {et~al.}(2013{\natexlab{a}})\citenamefont {Baroni}, \citenamefont
  {Navr\'{a}til},\ and\ \citenamefont {Quaglioni}}]{BaroniNQ13}%
  \BibitemOpen
  \bibfield  {author} {\bibinfo {author} {\bibfnamefont {S.}~\bibnamefont
  {Baroni}}, \bibinfo {author} {\bibfnamefont {P.}~\bibnamefont
  {Navr\'{a}til}}, \ and\ \bibinfo {author} {\bibfnamefont {S.}~\bibnamefont
  {Quaglioni}},\ }\href@noop {} {\bibfield  {journal} {\bibinfo  {journal}
  {Phys. Rev. Lett.}\ }\textbf {\bibinfo {volume} {110}},\ \bibinfo {pages}
  {022505} (\bibinfo {year} {2013}{\natexlab{a}})}\BibitemShut {NoStop}%
\bibitem [{\citenamefont {Baroni}\ \emph
  {et~al.}(2013{\natexlab{b}})\citenamefont {Baroni}, \citenamefont
  {Navr\'atil},\ and\ \citenamefont {Quaglioni}}]{PhysRevC.87.034326}%
  \BibitemOpen
  \bibfield  {author} {\bibinfo {author} {\bibfnamefont {S.}~\bibnamefont
  {Baroni}}, \bibinfo {author} {\bibfnamefont {P.}~\bibnamefont {Navr\'atil}},
  \ and\ \bibinfo {author} {\bibfnamefont {S.}~\bibnamefont {Quaglioni}},\
  }\href {\doibase 10.1103/PhysRevC.87.034326} {\bibfield  {journal} {\bibinfo
  {journal} {Phys. Rev. C}\ }\textbf {\bibinfo {volume} {87}},\ \bibinfo
  {pages} {034326} (\bibinfo {year} {2013}{\natexlab{b}})}\BibitemShut
  {NoStop}%
\bibitem [{\citenamefont {Hupin}\ \emph {et~al.}(2014)\citenamefont {Hupin},
  \citenamefont {Quaglioni},\ and\ \citenamefont
  {Navr\'atil}}]{PhysRevC.90.061601}%
  \BibitemOpen
  \bibfield  {author} {\bibinfo {author} {\bibfnamefont {G.}~\bibnamefont
  {Hupin}}, \bibinfo {author} {\bibfnamefont {S.}~\bibnamefont {Quaglioni}}, \
  and\ \bibinfo {author} {\bibfnamefont {P.}~\bibnamefont {Navr\'atil}},\
  }\href {\doibase 10.1103/PhysRevC.90.061601} {\bibfield  {journal} {\bibinfo
  {journal} {Phys. Rev. C}\ }\textbf {\bibinfo {volume} {90}},\ \bibinfo
  {pages} {061601} (\bibinfo {year} {2014})}\BibitemShut {NoStop}%
\bibitem [{\citenamefont {Calci}\ \emph {et~al.}(2016)\citenamefont {Calci},
  \citenamefont {Navr\'atil}, \citenamefont {Roth}, \citenamefont
  {Dohet-Eraly}, \citenamefont {Quaglioni},\ and\ \citenamefont
  {Hupin}}]{PhysRevLett.117.242501}%
  \BibitemOpen
  \bibfield  {author} {\bibinfo {author} {\bibfnamefont {A.}~\bibnamefont
  {Calci}}, \bibinfo {author} {\bibfnamefont {P.}~\bibnamefont {Navr\'atil}},
  \bibinfo {author} {\bibfnamefont {R.}~\bibnamefont {Roth}}, \bibinfo {author}
  {\bibfnamefont {J.}~\bibnamefont {Dohet-Eraly}}, \bibinfo {author}
  {\bibfnamefont {S.}~\bibnamefont {Quaglioni}}, \ and\ \bibinfo {author}
  {\bibfnamefont {G.}~\bibnamefont {Hupin}},\ }\href {\doibase
  10.1103/PhysRevLett.117.242501} {\bibfield  {journal} {\bibinfo  {journal}
  {Phys. Rev. Lett.}\ }\textbf {\bibinfo {volume} {117}},\ \bibinfo {pages}
  {242501} (\bibinfo {year} {2016})}\BibitemShut {NoStop}%
\bibitem [{\citenamefont {Hupin}\ \emph {et~al.}(2015)\citenamefont {Hupin},
  \citenamefont {Quaglioni},\ and\ \citenamefont
  {Navr\'atil}}]{PhysRevLett.114.212502}%
  \BibitemOpen
  \bibfield  {author} {\bibinfo {author} {\bibfnamefont {G.}~\bibnamefont
  {Hupin}}, \bibinfo {author} {\bibfnamefont {S.}~\bibnamefont {Quaglioni}}, \
  and\ \bibinfo {author} {\bibfnamefont {P.}~\bibnamefont {Navr\'atil}},\
  }\href {\doibase 10.1103/PhysRevLett.114.212502} {\bibfield  {journal}
  {\bibinfo  {journal} {Phys. Rev. Lett.}\ }\textbf {\bibinfo {volume} {114}},\
  \bibinfo {pages} {212502} (\bibinfo {year} {2015})}\BibitemShut {NoStop}%
\bibitem [{\citenamefont {Kravvaris}\ \emph {et~al.}(2020)\citenamefont
  {Kravvaris}, \citenamefont {Quaglioni}, \citenamefont {Hupin},\ and\
  \citenamefont {Navratil}}]{KravvarisQHN2020}%
  \BibitemOpen
  \bibfield  {author} {\bibinfo {author} {\bibfnamefont {K.}~\bibnamefont
  {Kravvaris}}, \bibinfo {author} {\bibfnamefont {S.}~\bibnamefont
  {Quaglioni}}, \bibinfo {author} {\bibfnamefont {G.}~\bibnamefont {Hupin}}, \
  and\ \bibinfo {author} {\bibfnamefont {P.}~\bibnamefont {Navratil}},\
  }\href@noop {} {\bibfield  {journal} {\bibinfo  {journal} {arXiv preprint
  arXiv:2012.00228}\ } (\bibinfo {year} {2020})}\BibitemShut {NoStop}%
\bibitem [{\citenamefont {Navrátil}\ \emph {et~al.}(2016)\citenamefont
  {Navrátil}, \citenamefont {Quaglioni}, \citenamefont {Hupin}, \citenamefont
  {Romero-Redondo},\ and\ \citenamefont {Calci}}]{1402-4896-91-5-053002}%
  \BibitemOpen
  \bibfield  {author} {\bibinfo {author} {\bibfnamefont {P.}~\bibnamefont
  {Navrátil}}, \bibinfo {author} {\bibfnamefont {S.}~\bibnamefont
  {Quaglioni}}, \bibinfo {author} {\bibfnamefont {G.}~\bibnamefont {Hupin}},
  \bibinfo {author} {\bibfnamefont {C.}~\bibnamefont {Romero-Redondo}}, \ and\
  \bibinfo {author} {\bibfnamefont {A.}~\bibnamefont {Calci}},\ }\href
  {http://stacks.iop.org/1402-4896/91/i=5/a=053002} {\bibfield  {journal}
  {\bibinfo  {journal} {Physica Scripta}\ }\textbf {\bibinfo {volume} {91}},\
  \bibinfo {pages} {053002} (\bibinfo {year} {2016})}\BibitemShut {NoStop}%
\bibitem [{\citenamefont {Jaganathen}\ \emph {et~al.}(2014)\citenamefont
  {Jaganathen}, \citenamefont {Michel},\ and\ \citenamefont
  {P{\l}oszajczak}}]{jaganathen_2014}%
  \BibitemOpen
  \bibfield  {author} {\bibinfo {author} {\bibfnamefont {Y.}~\bibnamefont
  {Jaganathen}}, \bibinfo {author} {\bibfnamefont {N.}~\bibnamefont {Michel}},
  \ and\ \bibinfo {author} {\bibfnamefont {M.}~\bibnamefont {P{\l}oszajczak}},\
  }\href@noop {} {\bibfield  {journal} {\bibinfo  {journal} {Phys. Rev. C}\
  }\textbf {\bibinfo {volume} {89}},\ \bibinfo {pages} {034624} (\bibinfo
  {year} {2014})}\BibitemShut {NoStop}%
\bibitem [{\citenamefont {Fossez}\ \emph {et~al.}(2015)\citenamefont {Fossez},
  \citenamefont {Michel}, \citenamefont {P{\l}oszajczak}, \citenamefont
  {Jaganathen},\ and\ \citenamefont {Betan}}]{fossez_2015}%
  \BibitemOpen
  \bibfield  {author} {\bibinfo {author} {\bibfnamefont {K.}~\bibnamefont
  {Fossez}}, \bibinfo {author} {\bibfnamefont {N.}~\bibnamefont {Michel}},
  \bibinfo {author} {\bibfnamefont {M.}~\bibnamefont {P{\l}oszajczak}},
  \bibinfo {author} {\bibfnamefont {Y.}~\bibnamefont {Jaganathen}}, \ and\
  \bibinfo {author} {\bibfnamefont {R.~I.}\ \bibnamefont {Betan}},\ }\href@noop
  {} {\bibfield  {journal} {\bibinfo  {journal} {Phys. Rev. C}\ }\textbf
  {\bibinfo {volume} {91}},\ \bibinfo {pages} {034609} (\bibinfo {year}
  {2015})}\BibitemShut {NoStop}%
\bibitem [{\citenamefont {Quaglioni}\ \emph {et~al.}(2013)\citenamefont
  {Quaglioni}, \citenamefont {Romero-Redondo},\ and\ \citenamefont
  {Navr\'atil}}]{PhysRevC.88.034320}%
  \BibitemOpen
  \bibfield  {author} {\bibinfo {author} {\bibfnamefont {S.}~\bibnamefont
  {Quaglioni}}, \bibinfo {author} {\bibfnamefont {C.}~\bibnamefont
  {Romero-Redondo}}, \ and\ \bibinfo {author} {\bibfnamefont {P.}~\bibnamefont
  {Navr\'atil}},\ }\href {\doibase 10.1103/PhysRevC.88.034320} {\bibfield
  {journal} {\bibinfo  {journal} {Phys. Rev. C}\ }\textbf {\bibinfo {volume}
  {88}},\ \bibinfo {pages} {034320} (\bibinfo {year} {2013})},\ \bibinfo {note}
  {[Erratum: Phys. Rev. C 94, 019902 (2016)]}\BibitemShut {NoStop}%
\bibitem [{\citenamefont {Descouvemont}\ and\ \citenamefont
  {Baye}(2010)}]{DescouvemontB10}%
  \BibitemOpen
  \bibfield  {author} {\bibinfo {author} {\bibfnamefont {P.}~\bibnamefont
  {Descouvemont}}\ and\ \bibinfo {author} {\bibfnamefont {D.}~\bibnamefont
  {Baye}},\ }\href@noop {} {\bibfield  {journal} {\bibinfo  {journal} {Rep.
  Prog. Phys.}\ }\textbf {\bibinfo {volume} {73}},\ \bibinfo {pages} {3}
  (\bibinfo {year} {2010})}\BibitemShut {NoStop}%
\bibitem [{\citenamefont {Descouvemont}(2016)}]{Descouvemont16}%
  \BibitemOpen
  \bibfield  {author} {\bibinfo {author} {\bibfnamefont {P.}~\bibnamefont
  {Descouvemont}},\ }\href@noop {} {\bibfield  {journal} {\bibinfo  {journal}
  {Comput. Phys. Commun.}\ }\textbf {\bibinfo {volume} {200}},\ \bibinfo
  {pages} {199} (\bibinfo {year} {2016})}\BibitemShut {NoStop}%
\bibitem [{\citenamefont {Casta{\~n}os}\ \emph {et~al.}(1988)\citenamefont
  {Casta{\~n}os}, \citenamefont {Draayer},\ and\ \citenamefont
  {Leschber}}]{CastanosDL88}%
  \BibitemOpen
  \bibfield  {author} {\bibinfo {author} {\bibfnamefont {O.}~\bibnamefont
  {Casta{\~n}os}}, \bibinfo {author} {\bibfnamefont {J.~P.}\ \bibnamefont
  {Draayer}}, \ and\ \bibinfo {author} {\bibfnamefont {Y.}~\bibnamefont
  {Leschber}},\ }\href@noop {} {\bibfield  {journal} {\bibinfo  {journal} {Z.
  Phys. A}\ }\textbf {\bibinfo {volume} {329}},\ \bibinfo {pages} {33}
  (\bibinfo {year} {1988})}\BibitemShut {NoStop}%
\bibitem [{\citenamefont {Mustonen}\ \emph {et~al.}(2018)\citenamefont
  {Mustonen}, \citenamefont {Gilbreth}, \citenamefont {Alhassid},\ and\
  \citenamefont {Bertsch}}]{MustonenGAB2018}%
  \BibitemOpen
  \bibfield  {author} {\bibinfo {author} {\bibfnamefont {M.~T.}\ \bibnamefont
  {Mustonen}}, \bibinfo {author} {\bibfnamefont {C.~N.}\ \bibnamefont
  {Gilbreth}}, \bibinfo {author} {\bibfnamefont {Y.}~\bibnamefont {Alhassid}},
  \ and\ \bibinfo {author} {\bibfnamefont {G.~F.}\ \bibnamefont {Bertsch}},\
  }\href {\doibase 10.1103/PhysRevC.98.034317} {\bibfield  {journal} {\bibinfo
  {journal} {Phys. Rev. C}\ }\textbf {\bibinfo {volume} {98}},\ \bibinfo
  {pages} {034317} (\bibinfo {year} {2018})}\BibitemShut {NoStop}%
\bibitem [{\citenamefont {Draayer}\ \emph {et~al.}(1989)\citenamefont
  {Draayer}, \citenamefont {Leschber}, \citenamefont {Park},\ and\
  \citenamefont {Lopez}}]{DraayerLPL89}%
  \BibitemOpen
  \bibfield  {author} {\bibinfo {author} {\bibfnamefont {J.~P.}\ \bibnamefont
  {Draayer}}, \bibinfo {author} {\bibfnamefont {Y.}~\bibnamefont {Leschber}},
  \bibinfo {author} {\bibfnamefont {S.~C.}\ \bibnamefont {Park}}, \ and\
  \bibinfo {author} {\bibfnamefont {R.}~\bibnamefont {Lopez}},\ }\href@noop {}
  {\bibfield  {journal} {\bibinfo  {journal} {Comput. Phys. Commun.}\ }\textbf
  {\bibinfo {volume} {56}},\ \bibinfo {pages} {279} (\bibinfo {year}
  {1989})}\BibitemShut {NoStop}%
\bibitem [{\citenamefont {Draayer}\ and\ \citenamefont
  {Akiyama}(1973)}]{DraayerSU3_1}%
  \BibitemOpen
  \bibfield  {author} {\bibinfo {author} {\bibfnamefont {J.~P.}\ \bibnamefont
  {Draayer}}\ and\ \bibinfo {author} {\bibfnamefont {Y.}~\bibnamefont
  {Akiyama}},\ }\href@noop {} {\bibfield  {journal} {\bibinfo  {journal} {J.
  Math. Phys.}\ }\textbf {\bibinfo {volume} {14}},\ \bibinfo {pages} {1904}
  (\bibinfo {year} {1973})}\BibitemShut {NoStop}%
\bibitem [{\citenamefont {Dytrych}\ \emph {et~al.}(2007)\citenamefont
  {Dytrych}, \citenamefont {Sviratcheva}, \citenamefont {Bahri}, \citenamefont
  {Draayer},\ and\ \citenamefont {Vary}}]{DytrychSBDV_PRL07}%
  \BibitemOpen
  \bibfield  {author} {\bibinfo {author} {\bibfnamefont {T.}~\bibnamefont
  {Dytrych}}, \bibinfo {author} {\bibfnamefont {K.~D.}\ \bibnamefont
  {Sviratcheva}}, \bibinfo {author} {\bibfnamefont {C.}~\bibnamefont {Bahri}},
  \bibinfo {author} {\bibfnamefont {J.~P.}\ \bibnamefont {Draayer}}, \ and\
  \bibinfo {author} {\bibfnamefont {J.~P.}\ \bibnamefont {Vary}},\ }\href@noop
  {} {\bibfield  {journal} {\bibinfo  {journal} {Phys. Rev. Lett.}\ }\textbf
  {\bibinfo {volume} {98}},\ \bibinfo {pages} {162503} (\bibinfo {year}
  {2007})}\BibitemShut {NoStop}%
\bibitem [{\citenamefont {Launey}\ \emph {et~al.}(2020)\citenamefont {Launey},
  \citenamefont {Dytrych}, \citenamefont {Sargsyan}, \citenamefont {Baker},\
  and\ \citenamefont {Draayer}}]{LauneyDSBD20}%
  \BibitemOpen
  \bibfield  {author} {\bibinfo {author} {\bibfnamefont {K.~D.}\ \bibnamefont
  {Launey}}, \bibinfo {author} {\bibfnamefont {T.}~\bibnamefont {Dytrych}},
  \bibinfo {author} {\bibfnamefont {G.~H.}\ \bibnamefont {Sargsyan}}, \bibinfo
  {author} {\bibfnamefont {R.~B.}\ \bibnamefont {Baker}}, \ and\ \bibinfo
  {author} {\bibfnamefont {J.~P.}\ \bibnamefont {Draayer}},\ }\href {\doibase
  10.1140/epjst/e2020-000178-3} {\bibfield  {journal} {\bibinfo  {journal}
  {Eur. Phys. J. Spec. Top.}\ }\textbf {\bibinfo {volume} {229}},\ \bibinfo
  {pages} {2429} (\bibinfo {year} {2020})}\BibitemShut {NoStop}%
\bibitem [{\citenamefont {Shirokov}\ \emph {et~al.}(2007)\citenamefont
  {Shirokov}, \citenamefont {Vary}, \citenamefont {Mazur},\ and\ \citenamefont
  {Weber}}]{ShirokovMZVW07}%
  \BibitemOpen
  \bibfield  {author} {\bibinfo {author} {\bibfnamefont {A.}~\bibnamefont
  {Shirokov}}, \bibinfo {author} {\bibfnamefont {J.}~\bibnamefont {Vary}},
  \bibinfo {author} {\bibfnamefont {A.}~\bibnamefont {Mazur}}, \ and\ \bibinfo
  {author} {\bibfnamefont {T.}~\bibnamefont {Weber}},\ }\href@noop {}
  {\bibfield  {journal} {\bibinfo  {journal} {Phys. Lett. B}\ }\textbf
  {\bibinfo {volume} {644}},\ \bibinfo {pages} {33} (\bibinfo {year}
  {2007})}\BibitemShut {NoStop}%
\bibitem [{\citenamefont {Baker}\ \emph
  {et~al.}(2020{\natexlab{b}})\citenamefont {Baker}, \citenamefont {Launey},
  \citenamefont {Bacca}, \citenamefont {Dinur},\ and\ \citenamefont
  {Dytrych}}]{baker_2020}%
  \BibitemOpen
  \bibfield  {author} {\bibinfo {author} {\bibfnamefont {R.~B.}\ \bibnamefont
  {Baker}}, \bibinfo {author} {\bibfnamefont {K.~D.}\ \bibnamefont {Launey}},
  \bibinfo {author} {\bibfnamefont {S.}~\bibnamefont {Bacca}}, \bibinfo
  {author} {\bibfnamefont {N.~N.}\ \bibnamefont {Dinur}}, \ and\ \bibinfo
  {author} {\bibfnamefont {T.}~\bibnamefont {Dytrych}},\ }\href@noop {}
  {\bibfield  {journal} {\bibinfo  {journal} {Phys. Rev. C}\ }\textbf {\bibinfo
  {volume} {102}},\ \bibinfo {pages} {014320} (\bibinfo {year}
  {2020}{\natexlab{b}})}\BibitemShut {NoStop}%
\bibitem [{\citenamefont {Bacca}\ \emph {et~al.}(2013)\citenamefont {Bacca},
  \citenamefont {Barnea}, \citenamefont {Hagen}, \citenamefont {Orlandini},\
  and\ \citenamefont {Papenbrock}}]{bacca:2013_prl}%
  \BibitemOpen
  \bibfield  {author} {\bibinfo {author} {\bibfnamefont {S.}~\bibnamefont
  {Bacca}}, \bibinfo {author} {\bibfnamefont {N.}~\bibnamefont {Barnea}},
  \bibinfo {author} {\bibfnamefont {G.}~\bibnamefont {Hagen}}, \bibinfo
  {author} {\bibfnamefont {G.}~\bibnamefont {Orlandini}}, \ and\ \bibinfo
  {author} {\bibfnamefont {T.}~\bibnamefont {Papenbrock}},\ }\href {\doibase
  10.1103/PhysRevLett.111.122502} {\bibfield  {journal} {\bibinfo  {journal}
  {Phys. Rev. Lett.}\ }\textbf {\bibinfo {volume} {111}},\ \bibinfo {pages}
  {122502} (\bibinfo {year} {2013})}\BibitemShut {NoStop}%
\bibitem [{\citenamefont {Ekstr{\"o}m}\ \emph {et~al.}(2013)\citenamefont
  {Ekstr{\"o}m}, \citenamefont {Baardsen}, \citenamefont {Forss{\'e}n},
  \citenamefont {Hagen}, \citenamefont {Hjorth-Jensen}, \citenamefont {Jansen},
  \citenamefont {Machleidt}, \citenamefont {Nazarewicz} \emph
  {et~al.}}]{Ekstrom13}%
  \BibitemOpen
  \bibfield  {author} {\bibinfo {author} {\bibfnamefont {A.}~\bibnamefont
  {Ekstr{\"o}m}}, \bibinfo {author} {\bibfnamefont {G.}~\bibnamefont
  {Baardsen}}, \bibinfo {author} {\bibfnamefont {C.}~\bibnamefont
  {Forss{\'e}n}}, \bibinfo {author} {\bibfnamefont {G.}~\bibnamefont {Hagen}},
  \bibinfo {author} {\bibfnamefont {M.}~\bibnamefont {Hjorth-Jensen}}, \bibinfo
  {author} {\bibfnamefont {G.~R.}\ \bibnamefont {Jansen}}, \bibinfo {author}
  {\bibfnamefont {R.}~\bibnamefont {Machleidt}}, \bibinfo {author}
  {\bibfnamefont {W.}~\bibnamefont {Nazarewicz}},  \emph {et~al.},\ }\href@noop
  {} {\bibfield  {journal} {\bibinfo  {journal} {Phys. Rev. Lett.}\ }\textbf
  {\bibinfo {volume} {110}},\ \bibinfo {pages} {192502} (\bibinfo {year}
  {2013})}\BibitemShut {NoStop}%
\bibitem [{\citenamefont {Burrows}\ \emph {et~al.}(2019)\citenamefont
  {Burrows}, \citenamefont {Elster}, \citenamefont {Weppner}, \citenamefont
  {Launey}, \citenamefont {Maris}, \citenamefont {Nogga},\ and\ \citenamefont
  {Popa}}]{BurrowsEWLMNP19}%
  \BibitemOpen
  \bibfield  {author} {\bibinfo {author} {\bibfnamefont {M.}~\bibnamefont
  {Burrows}}, \bibinfo {author} {\bibfnamefont {C.}~\bibnamefont {Elster}},
  \bibinfo {author} {\bibfnamefont {S.~P.}\ \bibnamefont {Weppner}}, \bibinfo
  {author} {\bibfnamefont {K.~D.}\ \bibnamefont {Launey}}, \bibinfo {author}
  {\bibfnamefont {P.}~\bibnamefont {Maris}}, \bibinfo {author} {\bibfnamefont
  {A.}~\bibnamefont {Nogga}}, \ and\ \bibinfo {author} {\bibfnamefont
  {G.}~\bibnamefont {Popa}},\ }\href {\doibase 10.1103/PhysRevC.99.044603}
  {\bibfield  {journal} {\bibinfo  {journal} {Phys. Rev. C}\ }\textbf {\bibinfo
  {volume} {99}},\ \bibinfo {pages} {044603} (\bibinfo {year}
  {2019})}\BibitemShut {NoStop}%
\bibitem [{\citenamefont {Navr\'atil}\ \emph {et~al.}(2010)\citenamefont
  {Navr\'atil}, \citenamefont {Roth},\ and\ \citenamefont
  {Quaglioni}}]{PhysRevC.82.034609}%
  \BibitemOpen
  \bibfield  {author} {\bibinfo {author} {\bibfnamefont {P.}~\bibnamefont
  {Navr\'atil}}, \bibinfo {author} {\bibfnamefont {R.}~\bibnamefont {Roth}}, \
  and\ \bibinfo {author} {\bibfnamefont {S.}~\bibnamefont {Quaglioni}},\ }\href
  {\doibase 10.1103/PhysRevC.82.034609} {\bibfield  {journal} {\bibinfo
  {journal} {Phys. Rev. C}\ }\textbf {\bibinfo {volume} {82}},\ \bibinfo
  {pages} {034609} (\bibinfo {year} {2010})}\BibitemShut {NoStop}%
\bibitem [{\citenamefont {Ekstr\"om}\ \emph {et~al.}(2015)\citenamefont
  {Ekstr\"om}, \citenamefont {Jansen}, \citenamefont {Wendt}, \citenamefont
  {Hagen}, \citenamefont {Papenbrock}, \citenamefont {Carlsson}, \citenamefont
  {Forss\'en}, \citenamefont {Hjorth-Jensen}, \citenamefont {Navr\'atil},\ and\
  \citenamefont {Nazarewicz}}]{PhysRevC.91.051301}%
  \BibitemOpen
  \bibfield  {author} {\bibinfo {author} {\bibfnamefont {A.}~\bibnamefont
  {Ekstr\"om}}, \bibinfo {author} {\bibfnamefont {G.~R.}\ \bibnamefont
  {Jansen}}, \bibinfo {author} {\bibfnamefont {K.~A.}\ \bibnamefont {Wendt}},
  \bibinfo {author} {\bibfnamefont {G.}~\bibnamefont {Hagen}}, \bibinfo
  {author} {\bibfnamefont {T.}~\bibnamefont {Papenbrock}}, \bibinfo {author}
  {\bibfnamefont {B.~D.}\ \bibnamefont {Carlsson}}, \bibinfo {author}
  {\bibfnamefont {C.}~\bibnamefont {Forss\'en}}, \bibinfo {author}
  {\bibfnamefont {M.}~\bibnamefont {Hjorth-Jensen}}, \bibinfo {author}
  {\bibfnamefont {P.}~\bibnamefont {Navr\'atil}}, \ and\ \bibinfo {author}
  {\bibfnamefont {W.}~\bibnamefont {Nazarewicz}},\ }\href {\doibase
  10.1103/PhysRevC.91.051301} {\bibfield  {journal} {\bibinfo  {journal} {Phys.
  Rev. C}\ }\textbf {\bibinfo {volume} {91}},\ \bibinfo {pages} {051301}
  (\bibinfo {year} {2015})}\BibitemShut {NoStop}%
\bibitem [{\citenamefont {Launey}\ \emph {et~al.}(2012)\citenamefont {Launey},
  \citenamefont {Dytrych},\ and\ \citenamefont {Draayer}}]{LauneyDD12}%
  \BibitemOpen
  \bibfield  {author} {\bibinfo {author} {\bibfnamefont {K.~D.}\ \bibnamefont
  {Launey}}, \bibinfo {author} {\bibfnamefont {T.}~\bibnamefont {Dytrych}}, \
  and\ \bibinfo {author} {\bibfnamefont {J.~P.}\ \bibnamefont {Draayer}},\
  }\href@noop {} {\bibfield  {journal} {\bibinfo  {journal} {Phys. Rev. C}\
  }\textbf {\bibinfo {volume} {84}},\ \bibinfo {pages} {044003} (\bibinfo
  {year} {2012})}\BibitemShut {NoStop}%
\end{thebibliography}%

\end{document}